\documentclass[%
 reprint,
nobibnotes,
 amsmath,amssymb,
 aps,
]{revtex4-2}

\usepackage[utf8]{inputenc}
\usepackage{graphicx}				
\usepackage{dcolumn}				
\usepackage{bm}						
\usepackage{hyperref}
\usepackage{geometry}
\usepackage[dvipsnames]{xcolor}
\usepackage{float}
\usepackage{multirow}
\usepackage{mathtools}
\usepackage{overpic}
\usepackage{enumitem}
\usepackage{titlesec}
\usepackage{booktabs}
\usepackage{algorithm}
\usepackage{algpseudocode}
\usepackage{csquotes}
\MakeOuterQuote{"}

\hypersetup{colorlinks=true,allcolors=blue}

\setcitestyle{authoryear}
\setcitestyle{round}

\setlist{itemsep=-5pt, topsep=0pt}
\titleformat{\paragraph}[runin]
  {\bigskip\normalfont\bfseries} 
  {} 
  {0pt} 
  {} 
\titlespacing*{\paragraph}{0pt}{\parskip}{1em}

\begin{document}



\title{Spatial Patterning and Selection: \\ How the Environment Shapes Molecular Complexity}

\author{Alexandre Champagne-Ruel}
\affiliation{
	Département de physique, 
	Université de Montréal 
	H2V 0B3, 
	Canada
}
\affiliation{
	Biodesign Institute, 
	Arizona State University, 
	Tempe, 
	Arizona 85281, 
	USA
}

\author{Christopher P. Kempes}
\affiliation{
    Santa Fe Institute, 
    Santa Fe, 
    New Mexico 87501, 
    USA
}

\author{Cole Mathis}
\email{cole.mathis@asu.edu}
\affiliation{
	Biodesign Institute, 
	Arizona State University, 
	Tempe, 
	Arizona 85281, 
	USA
}
\affiliation{
	School of Complex Adaptive Systems, 
	Arizona State University, 
	Tempe, 
	Arizona 85281, 
	USA
}

\date{September 4, 2025}


\begin{abstract}
	Assembly theory predicts that a distinguishing signature of life is its ability to produce complex molecules in abundance, opening new possibilities for life detection. 
	Experimental validation of this approach has so far relied on abiotic controls like meteoritic material, or simple, well-mixed chemical systems. 
	However, decades of research in self-organization have shown that spatial patterning can foster dynamical self-organization. 
	This raises the possibility that systems with nontrivial spatial patterns might promote abiotic formation of molecules with higher than expected assembly indices, potentially leading to false positives in life detection approaches based on assembly theory. 
	To explore this, we used a model of artificial chemistry to investigate how spatial organization can influence the development of molecular assembly indices. 
	Our findings reveal that transport factors, such as diffusion, significantly affect the distribution of chemical species within a system.
	Additionally, system topology critically impacts the distribution of assembly indices: while it does not enable arbitrary complexity, ordered lattices can shift the threshold for abiotic chemistry upward.
	Moreover, we demonstrate that diffusion can impede the formation and detection of these high assembly index molecules, bearing important implications for life detection experiments and astrobiological missions.
\end{abstract}


\maketitle



\section{Introduction}
\label{sec:intro}

Despite significant scientific advances, two fundamental questions about life remain unanswered:
How did life emerge, and how can we reliably distinguish life from non-life?

Recent research in assembly theory presents empirical evidence suggesting that only living systems can generate sufficiently complex molecules, quantifiable by their \textit{assembly index} (AI), in detectable quantities \citep{marshall_identifying_2021}. 
However, the abiotic controls referenced in this study are limited to two sources: (i) meteoritic material and (ii) laboratory experiments conducted in well-mixed systems. 
Notably, spatial heterogeneity, prevalent in diverse natural and artificial environments such as porous materials or mineral surfaces, has been demonstrated to catalyze dynamic self-organization, often described as "complexity".

This raises a critical question: does spatial heterogeneity enhance or impede the formation of high assembly index molecules? 
To address this, we have developed a numerical model aimed at testing the hypothesis that spatial patterning can induce the formation of higher molecular assembly indices in the absence of living systems. 
Our findings aim to inform future empirical research, particularly in minimizing false positives in life detection and fostering conditions conducive to the emergence of \textit{de novo} life.

Assembly theory (AT) predicts living systems are uniquely able to produce high assembly index molecules in abundance \cite{marshall_identifying_2021,sharma_assembly_2023}. 
AI is a central measure in AT as a quantitative measure of complexity, defined as the minimal number of joining operations required to construct an object from basic components. 
This metric allows for the structural comparison of diverse objects made from common building blocks. 
For covalently bonded molecules, AI (also known as molecular assembly (MA) in the context of molecules) is observable via measurements based on mass spectrometry and as well as spectroscopic experiments, enabling its empirical determination~\citep{jirasek_multimodal_2023,marshall_identifying_2021}. 
By combining this measure of structural complexity with the copy number of those molecules, AT evaluates the extent of selective processes~—~Darwinian or otherwise~—~in its formation~\citep{sharma_assembly_2023}. 



Empirical tests of the theory are consistent with the prediction that only living systems are capable of generating large quantities of high AI molecules, reinforcing the robustness of life detection experiments based on AI against false positives \citep{marshall_identifying_2021}. 
Nonetheless, isolating completely abiotic samples to further test this hypothesis proves challenging. 
This difficulty arises because virtually every Earth environment has experienced some form of biological influence, and even ancient materials often display signatures potentially linked to life ~\citep{dragone2021exploring, ratliff2023vacant, bell2015}. 
In their 2021 study, Marshall et al. utilized meteoritic material as an abiotic control alongside products from canonical prebiotic experiments. 
However, replicating the diverse chemical conditions characteristic of planetary surfaces in laboratory settings remains a significant challenge~\citep{fortney2019needlaboratorymeasurementsab, fayolle2020criticallaboratorystudiesadvance}.
A critical factor contributing to this challenge is the inherent spatial heterogeneity observed in natural systems, which contrasts with the uniform conditions typical of laboratory and numerical experiments, including the well-mixed prebiotic controls used in Marshall et al. (2021).


Rocks, inherently porous materials, feature numerous micro-compartments that may drive nontrivial physio-chemical processes \citep{matreux_heat_2021, matreux_heat_2024}.
These compartments create spatial heterogeneity, or patterning. 
Complexity science has identified spatial patterning as a fundamental mechanism responsible for fostering dynamic self-organization. 
Historically, the interaction between non-linear dynamics and diffusion has been known to produce spatial patterns, notably in biological systems \citep{turing_chemical_1952}. 
Furthermore, computational studies suggest spiral waves may aid prebiotic evolution by impeding parasites \citep{boerlijst_spiral_1991}, and spatial self-organization can support RNA-like replicators \citep{colizzi_parasites_2016}. 
This phenomenon has been extensively examined in computational biology and game theory, revealing how diverse processes shape spatial patterns \citep{nowak_spatial_1993, lindgren_evolutionary_1994}, with significant implications for origins-of-life research \citep{mizuuchi_sustainable_2018, champagne-ruel_mutation_2022}.

Laboratory investigations of mineral surfaces have unveiled their influence on polymerization.
Specifically, serpentine environments have been observed to favour glycine- and phenylalanine-rich fragments, leading to the formation of longer oligomers \citep{asche_evidence_2024}.
Similarly, analyses of asteroid samples, such as those from Ryugu, underscore the role of mineral surfaces in the spatial distribution of organic matter \citep{hashiguchi_spatial_2023}.
These laboratory studies reveal that minerals contribute both catalytic activity and the formation of nontrivial micro-compartments, each potentially affecting the resultant product distribution.
However, disentangling the distinct impacts of catalysis and micro-compartmentalization remains challenging and the isolated effect of spatial patterning on product distribution is not well understood.
Recent findings suggest that geo-compartments, even in the absence of catalytic activity, can selectively segregate prebiotic building blocks, a process akin to thermophoresis \citep{matreux_heat_2024}.


We hypothesize that spatial patterning alone may alter the distribution of molecular assembly indices, independently of biological processes. 
Given the importance of porous environments in prebiotic chemistry, we systematically investigate how spatial patterning and fluid transport properties affect the emergence of molecular complexity. 
By varying inflow and diffusion parameters in our model, we simulate conditions resembling diverse environments~—~ranging from hydrothermal vents (e.g. high inflow, low diffusion) \citep{rasmussen_nanoparticulate_2024} to tide pools (e.g. low inflow, high diffusion). 
Our approach seeks general insights rather than focusing on specific environments, aiming to delineate conditions conducive to abiotic synthesis of higher assembly index molecules, thereby false-positive risks in life detection efforts by informing appropriate thresholds for confident life detection.

\medskip

We employ a chemical model that simplifies the intricacy of real chemical systems to basic operations of integer addition and subtraction.
This means that we represent all molecules as integers, and reactions between them as either addition of two integers, or decomposition of an integer into smaller ones (see Table~\ref{tab:reactions}).
\cite{liu_mathematical_2018} used a similar approach to model reactions such as the citric acid cycle and the formose reaction, revealing that certain chemical systems exhibit collective catalysis through intermediates, while others demonstrate full self-replication. 
These results suggest that such catalytic behaviors commonly emerge in random artificial chemistries, consistent with previous studies \citep{banzhaf_artificial_2015}.

This model abstracts two fundamental characteristics inherent in real chemical systems: (i) composition and (ii) open-endedness.
Composition refers to the ability of objects to combine and form new entities, with decomposition serving as the reverse process.
Open-endedness signifies the system's capacity to generate an unlimited number of unique objects~—~analogous to the infinite range of positive integers and the boundless diversity of chemical compounds.
We apply AT to bridge computational (\textit{in silico}) models and experimental (lab-based) research. 
Specifically, we use AI as our complexity metric for individual molecules because of its direct relevance to laboratory and field measurements. 
Importantly, real chemical space is far richer than the abstract integer-based model; for example, many distinct molecules can share the same number of atoms or bonds.
This means that the assembly indices computed here and the assembly indices of covalent molecules may represent distinct scales, and must be interpreted carefully in that context.

The spatial organization of porous media can be effectively represented as a network of nanopores, enabling the application of network theory \citep{albert_statistical_2002, barabasi_network_2016}. 
Within this framework, we model the interconnections between nanopores as directed graphs with nodes (vertices) connected by edges (links). 
We take this to physically represent nanopores (nodes) connected via channels (edges) within porous media.
We specifically focus on regular lattice structures to systematically explore the influence of topology on molecular assembly indices.

Studying how environmental conditions impact biosignatures is essential for optimizing life detection methods and understanding life's emergence. 
A critical question is whether spatial heterogeneity alone enhances or inhibits the formation of higher assembly index molecules. 
Our numerical approach specifically investigates this relationship.
Our results indicate that environmental topology and transport dynamics, particularly diffusion, are pivotal in shaping molecular complexity within abiotic systems.
Specifically, we observed that structured network topologies, such as regular lattices, promote the formation of higher assembly index molecules, whereas randomized configurations tend to inhibit them.

These findings may help refine life detection strategies by identifying abiotic conditions capable of modulating the distribution of assembly indices in the absence of life~—~knowledge that is crucial for appropriately parameterizing life detection methods. 
Furthermore, characterizing the processes by which the complexity of abiotic systems can increase may also yield insights into the \textit{de novo} synthesis of life.


\section{Methods}
\label{sec:methods}


\paragraph{Integer Calculus.} 

\begin{figure*}[t]
    \centering
    \begin{minipage}{0.49\textwidth}
        \centering
        \hspace{2em}
        \begin{overpic}[width=0.79\textwidth]{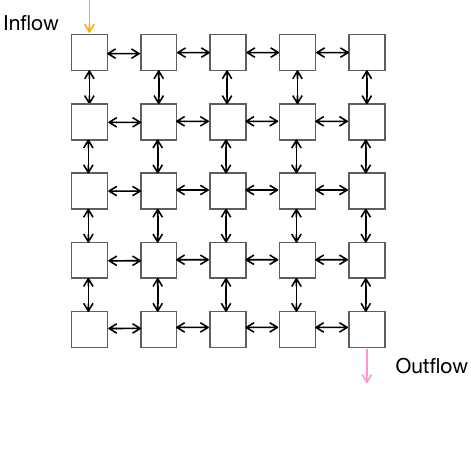}
            \put(-2, 41){\textbf{(A)}}
        \end{overpic}
        \vspace{1em}
        \begin{overpic}[scale=0.75]{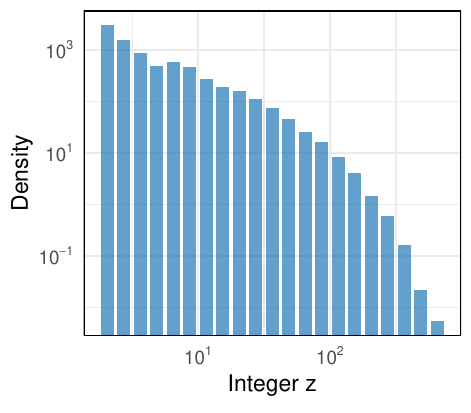}
            \put(0, 80){\textbf{(B)}}
        \end{overpic}
    \end{minipage}
    \hfill
    \begin{minipage}{0.49\textwidth}
        \centering
        \begin{overpic}[width=\textwidth]{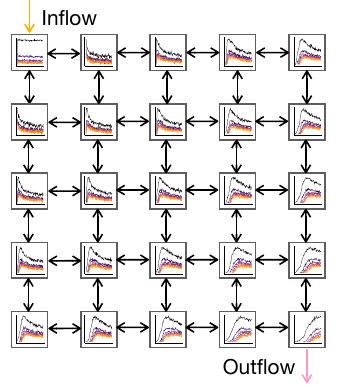}
            \put(-7, 91){\textbf{(C)}}
        \end{overpic}
    \end{minipage}
    
    \caption{
        \textbf{(A)} Diagram depicting the model’s chemostats (reactors), which contain integers capable of undergoing five transformation types: inflow, diffusion, outflow, synthesis (forward reactions), and decomposition (backward reactions) (refer to Table~\ref{tab:reactions}). 
        \textbf{(B)} Representative simulation showcasing the final molecular distribution across the whole network of reactors.
        \textbf{(C)} Simulation framework comprising an ensemble of reactors with uniform volume, temperature, and pressure, interconnected in a specific topology (e.g., regular lattice). 
        This setup allows bidirectional diffusion between neighboring reactors. 
        Numerical integration employs the $\tau$-leaping method to generate population time-series for each reactor, emphasizing dynamic interactions over detailed chemical properties.
    }
    \label{fig:model}
\end{figure*}

The model we developed comprises a set of reactors (chemostats) containing integers (Fig.~\ref{fig:model}A), capable of undergoing five types of transformations: inflow, forward reactions (synthesis), backward reactions (decomposition), diffusion (movement to neighbouring reactors), and outflow (out of the system), as detailed in Table~\ref{tab:reactions}.
Inflow, represented as a zeroth-order reaction, occurs at a rate $I$ in specific reactors (refer to results below) and is measured in units of molecules per unit time.

Backward (destructive) reactions, diffusion and outflow, are modelled as first order reactions, and have corresponding rate constants $k_b$, $k_d$ and $k_o$, respectively.
To simplify the parameter space, the backward reaction rate constant has been normalized with $k_b \equiv 1$ throughout all simulations, and the diffusion coefficient assumed equal to the outflow rate constant by setting $k_d = k_o$, reducing the number of free parameters.
Forward (constructive) reactions are second-order processes driven by pairwise interactions, occurring at a rate constant of $k_f$ within local spatial volume.
All other reaction rate parameters are held constant during simulations to reflect steady-state kinetics, aligning with conditions typical of experiments conducted under stable environmental conditions \citep{rodriguez-garcia_formation_2015}.

Table~\ref{tab:reactions} provides a summary of the reaction rates utilized in the model.
To minimize bias introduced in the resulting chemistry, we maintain a constant backward reaction rate across all molecules, independent of molecule length (i.e., integer value).
While alternative approaches could consider that larger molecules might break more readily due to an increased number of bonds or, conversely, be more resistant to degradation due to enhanced stability, our intermediate position seeks to balance these perspectives without introducing variability.
This approach may yield molecules of higher mass and therefore possibly of high assembly indices, and since we are primarily concerned with minimizing false positives this is a more conservative simplification.

By applying uniform constants for both forward and backward reaction rates across all pairwise interactions and degradation reactions, we aim to maximize the diversity within our model.
In contrast, natural chemical systems exhibit reaction rate constants that are highly dependent on the specific components involved, often differing by orders of magnitude.
These kinetic disparities in real chemistry tend to restrict the ensemble of accessible molecules within a given environment.
Nevertheless, our integer-based abstract model is inherently more constrained than the vast molecular space found in natural systems, as there can be numerous molecules with $N$ bonds but only a single corresponding positive integer of size $N$.
Consequently, we anticipate the ensemble of accessible molecules in our simulations to be considerably smaller compared to real chemical environments.


\begin{table}[t]
    \centering
    \begin{tabular}{lcc}
        \toprule
        \textbf{Reaction Type} & \textbf{Reaction} & \textbf{Propensity} \\
        \midrule
        Inflow             & $\varnothing \xrightarrow{I} A$               & $I$ \\
        Diffusion          & $C_i \xleftrightarrow{k_d} C_j$              & $k_d [C_i]$ \\
        Outflow            & $C_i \xrightarrow{k_o} \varnothing$          & $k_o [C_i]$ \\
        Forward reaction   & $A + B \xrightarrow{k_f} C$                  & $k_f [A][B]$ \\
        Backward reaction  & $C \xrightarrow{k_b} A^{\prime} + B$         & $k_b [C]$ \\
        \bottomrule
    \end{tabular}
    \caption{
        Overview of reactions in the artificial chemistry system. Designated reactors receive an inflow of $I$ molecules per unit time.
        Molecules diffuse to neighboring chemostats at a rate of $k_d$, while outflow occurs at a rate of $k_o$.
        Constructive (forward) reactions proceed at a rate of $k_f$, and destructive (backward) reactions occur at a rate of $k_b$, independent of molecule size.
    }
    \label{tab:reactions}
\end{table}


\paragraph{Numerical Integration with the Gillespie Algorithm.}

The model's numerical integration employs a stochastic methodology, incorporating both the exact Stochastic Simulation Algorithm (SSA) \citep{gillespie_exact_1977} and the $\tau$-leaping method \citep{gillespie_approximate_2001}.
Unless specified otherwise, the results presented herein are derived from the $\tau$-leaping scheme.

Let $k_i$ denote one reaction rate.
The corresponding reaction propensity, $\pi_i$, is calculated by multiplying this reaction rate by the number of combinatorial possibilities that facilitate the reaction.
Specifically, for reactions driven by pair-wise collisions in a mixture of $n_i$ molecules within reactor $i$, this relationship can be articulated as
\begin{equation}
	\pi_f = k_f \cdot n_i(n_i-1).
\end{equation}
Note that this includes the terms for the rate constant and the volume in the second order reaction, for simplicity.
While the original SSA yields precise outcomes, its computational demands become prohibitive with a large number of reactions.

To address this, we adopt the "$\tau$-leaping" technique, which approximates the number of reactions within a time interval $\text{d}t$ using a Poisson distribution with a mean of $\pi_i \cdot \text{d}t$ {\citep{gillespie_approximate_2001}}:
\begin{equation}
n_i \approx \mathcal{P}(\pi_i \cdot \text{d}t).
\end{equation}
 The subsequent reaction time is therefore calculated as:
\begin{equation}
    \tau = \frac{-\log R}{\sum_{i=1}^N \pi_i}
\end{equation}
 where $R$ is a pseudo-random number within the interval $[0,1]$,
leading to significantly reduced calculation times and achieving speed-ups of one to two orders of magnitude.


\paragraph{Modelling Ensembles of Reactors.}

In our simulations, multiple reactors are interconnected following a fixed topology over time, permitting bidirectional diffusion (Fig.~\ref{fig:model}C).
We assume uniform conditions across all reactors, maintaining the same volume, temperature, and pressure.
Consequently, our reactions adhere only to the principle of mass action, while other  typical chemical properties, such as kinetic rates or thermodynamic stability influenced by internal structures have been abstracted away.



\paragraph{Calculation of the Assembly Index.}


We represent basic chemical units as integers and approximate the assembly index of an integer $n$ using the minimal length of all addition chains that terminate at $n$ \citep{marshall_formalising_2022}.
This serves as a conceptual lower bound for the assembly index of a molecule with $n$ bonds, given that the assembly space of covalently bonded molecules can be mapped onto the assembly space of integer addition (see \citep{marshall_formalising_2022} for details).

\begin{figure}[t]
    \centering
    \includegraphics[width=0.49\textwidth]{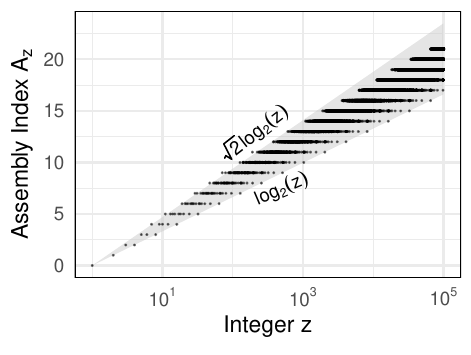}
    \caption{
        Computation of the assembly index for integers using an optimized algorithm designed for minimal-length addition chains. 
        The plot showcases the logarithmic range of the assembly index, bounded by $\log_2(z) < A_z \lesssim \sqrt{2} \log_2(z)$. 
        This finding establishes a conceptual lower limit for the assembly index of molecules with $n$ bonds, drawing parallels between the assembly spaces of covalently bonded molecules and integer addition.
    }
    \label{fig:integers-assembly}
\end{figure}

To determine the assembly index for integers, we utilize a numerical implementation of an efficient algorithm designed to generate minimal-length addition chains \citep{thurber_efficient_1999}.
Our findings indicate that, for integers, the assembly index is logarithmically bounded as $\log_2(z) < A_z \lesssim \sqrt{2}\log_2(z)$ (Fig.~\ref{fig:integers-assembly}).
Although we can compute assembly indices within this framework, mapping these results to empirical systems is non-trivial.
This difficulty arises partly because, for every positive integer $N$, there exists a super-exponential growth in the number of covalently bonded molecules with $N$ bonds.
As a result, the dynamics of our model are highly degenerate when compared to real chemical dynamics.
While assembly theory gives us a consistent framework within which to evaluate these scales, mapping from one scale to another is a topic for future work.

\section{Results}
\label{sec:results}


\paragraph{Varying Diffusion and Inflow.}
To assess the parameters defining porous materials, we first investigate how diffusion influences the distribution of molecular species.
Simulations were performed with a statistical ensemble of the previously detailed artificial chemistry models to monitor changes in molecular species distribution.
Our approach employs the $\tau$-leaping algorithm to evolve systems arranged in a regular lattice of $N=25$ reactors, allowing for bidirectional diffusion.
The model evolves with a time step of $d\text{t} = 10^{-3}$ until reaching $T = 10^5$.

\begin{figure*}[t]
    \vspace{2em}
    \centering

    \begin{overpic}[width=0.49\textwidth]{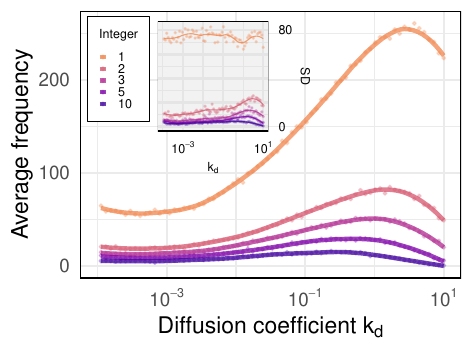}
        \put(0, 75){\textbf{(A)}}
    \end{overpic}
    \begin{overpic}[width=0.49\textwidth]{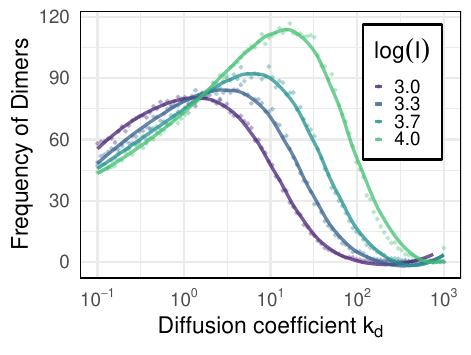}
        \put(0, 75){\textbf{(B)}}
    \end{overpic}

    \vspace{1em}

    \begin{overpic}[width=0.49\textwidth]{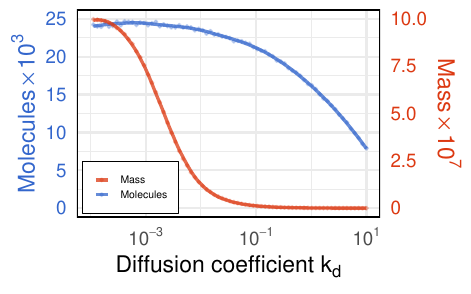}
        \put(-5, 65){\textbf{(C)}}
    \end{overpic}
    \hspace{0.025\textwidth}
    \begin{overpic}[width=0.40\textwidth]{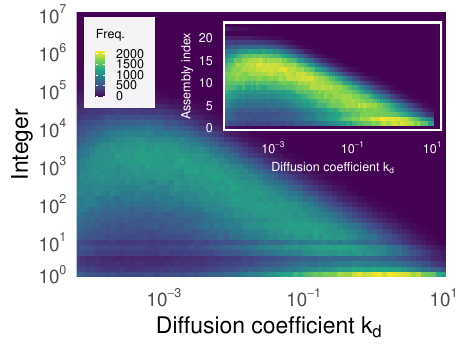}
        \put(-13, 80){\textbf{(D)}}
    \end{overpic}

    \caption{
        \textbf{(A)} Final populations of integers $z \in \{1, \cdots, 10\}$ at $t = 10^5$ as a function of the diffusion parameter $k_d$.
        Each point is the mean across all $N = 25$ reactors, with the inset showing standard deviations.
        Three diffusion regimes emerge: 
        (i) low diffusion ($\log k_d \lesssim -3$) with local accumulation and non-equilibrated systems; 
        (ii) intermediate diffusion ($-3 \lesssim \log k_d \lesssim -1$) where complexity is minimized due to lower interaction rates; and 
        (iii) high diffusion ($\log k_d \gtrsim -1$) resembling well-mixed systems with increased complexity.
        \textbf{(B)} Dimer populations ($z=2$) as a function of diffusion and inflow ($\log I \in \{3, \cdots, 4\}$).
        Higher inflows raise dimer concentrations and shift the $k_d$ value for maximal abundance, due to increased second-order reaction rates.
        \textbf{(C)} Number of molecules and system mass across diffusion regimes.
        At low $k_d$, mass is stable, but molecule count rises due to sparse constructive events.
        As diffusion increases, the mass drops faster than molecule number, suggesting declining complexity.
        \textbf{(D)} Integer frequency distributions for $z \in \{1, \cdots, 10^7\}$ over 50 log-spaced bins and various $k_d$ values.
        Heatmaps reveal population shifts toward higher integers with increasing diffusion, then a decline at high $k_d$.
    }
    \label{fig:varying-diffusion}
\end{figure*}

Figure~\ref{fig:varying-diffusion} illustrates a statistical ensemble of $N=100$ simulations, sampling diffusion rates across the range $\log k_d \in \{-4, \cdots, 1\}$.
Reactors are initially empty. One designated inflow reactor receives $I=10^3$ monomers per time step, while an outflow reactor is emptied at a rate equal to the diffusion rate. The inflow and outflow reactors are placed at the maximal possible distance from each other on a 2D lattice under a von Neumann connection scheme, in which each site connects to its four cardinal neighbors~(see Figure~\ref{fig:model}C)~\citep{toffoli1987cellular}. The constructive reaction rate is set to $k_f = 10^{-3}$.

We first show in Figure~\ref{fig:varying-diffusion}A the average final populations ($t = 10^5$) of integers $z \in \{1, \cdots, 10\}$, computed across all reactors in each simulation, as a function of the diffusion parameter $k_d$.
Three distinct regimes emerge.
At low diffusion rates ($\log k_d \lesssim -3$), the system has not yet reached steady state due to limited molecular transport; most molecules remain in the inflow reactor, which behaves as a well-mixed system.
As diffusion increases ($-3 \lesssim \log k_d \lesssim -1$), molecules spread across reactors, but reduced local concentrations still hinder reactions.
At higher diffusion values ($\log k_d \gtrsim -1$), all reactors become populated to varying extents.
The system approaches homogeneity again, allowing the population of these species to rise~—~until diffusion becomes so fast ($\log k_d \sim 1$) that molecules are rapidly lost through outflow, causing another drop in population.
The inset of Figure~\ref{fig:varying-diffusion}A presents the standard deviation in population across reactors.
It peaks at high diffusion due to inflow accumulation, reflecting rapid turnover and spatial imbalance.

Our analysis also examined the impact of varying inflow rates within these higher diffusion regimes.
This approach allowed us to differentiate between the effects of transport and inflow rates.
Figure~\ref{fig:varying-diffusion}B illustrates the dimer population ($z=2$) across different inflow rates, specifically for $\log I \in \{3, \cdots, 4\}$, spanning intermediate and high diffusion conditions.

As inflow increases, the distribution of the dimer population shifts, peaking at high values of the diffusion coefficient, with the maximum reaching higher levels. 
In the high diffusion regime ($\log k_d \gtrsim 0$), greater inflows result in increased dimer populations, with this increase being correlated with the rise in inflow. 
However, when diffusion exceeds a critical value ($\log k_d \gtrsim 2$), dimer populations decrease again until they reach zero. 
In this regime, transport dynamics dominate over other processes, completely preventing the formation of larger molecules.

Figure~\ref{fig:varying-diffusion}C illustrates the variations in both the number of molecules and mass across different diffusion regimes.
At low diffusion regimes, where the majority of molecules remain localized in the inflow reactor, mass remains unaffected by the diffusion parameter.
This is due to the limited diffusion, preventing molecules from reaching the outflow reactor.
In this regime, the number of molecules increases because as they disperse through the reactors, their likelihood of undergoing constructive reactions diminishes, leading to a higher overall quantity of smaller molecules.
As diffusion continues to rise, molecules gradually populate all reactors.
In this phase, mass decreases more rapidly than the number of molecules, suggesting a reduction in high integer species.
This occurs because molecules are transported swiftly through the system, leaving insufficient time for reactions necessary to form larger molecules.

Figure~\ref{fig:varying-diffusion}D supports these intuitions by illustrating the frequency distribution of integers $z \in \{1, \cdots, 10^7\}$, divided into 50 logarithmically spaced bins across varying diffusion coefficients.
In the heatmap, lighter colors denote higher population densities.
At the lowest diffusion levels, the majority of molecules correspond to lower integer values.
As diffusion increases, the distribution's mean shifts towards higher integers, only to decline again when $k_d$ reaches very high values.

The inset in Figure~\ref{fig:varying-diffusion}D illustrates the distribution of assembly indices, now presented on a linear vertical scale, in relation to the diffusion coefficient. As AI is proportional to the logarithm of the integer value, specifically $a_z \propto \log_2 (z)$, the overall trends for both the integer and AI distributions are comparable.


\paragraph{Assembly.}
A central metric in Assembly Theory for evaluating the complexity of a physical system is the Assembly ($A$), which \cite{sharma_assembly_2023} define as:
\begin{equation}
A = \sum_{i=1}^{N} e^{a_i} \left( \frac{n_i - 1}{N_{\mathrm{T}}} \right)
\end{equation}
where $a_i$ and $n_i$ are the Assembly Index and copy number of object $i$, $N$ is the number of unique objects in the system, and $N_T$ is the total (non-unique) number of objects.
We used $A$ to track the temporal evolution of system-wide complexity in three representative simulations across low, intermediate, and high diffusion regimes.
We also computed the difference in complexity between the inflow and outflow, referred to as $\Delta A$.

\begin{figure*}[t]
    \centering
    \begin{overpic}[width=0.49\textwidth]{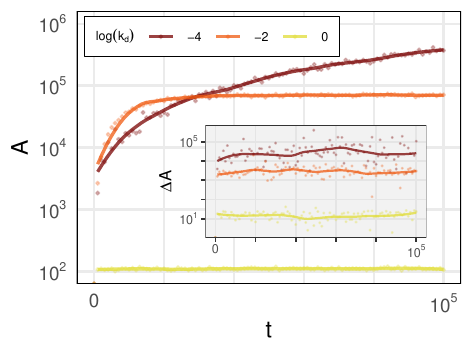}
        \put(0, 75){\textbf{(A)}}
    \end{overpic}
    \begin{overpic}[width=0.49\textwidth]{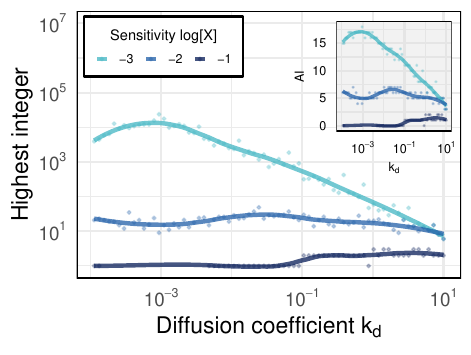}
        \put(0, 75){\textbf{(B)}}
    \end{overpic}
    \caption{
        \textbf{(A)} Temporal evolution of total Assembly ($A$) and the Assembly difference between inflow and outflow ($\Delta A$) across three diffusion regimes.
        The red curve corresponds to low diffusion ($\log k_d \sim -4$), showing a rapid initial increase in Assembly followed by a slow plateau.
        The orange curve represents intermediate diffusion ($\log k_d \sim -2$), where Assembly stabilizes sooner due to spatial dispersion limiting constructive reactions.
        The yellow curve depicts high diffusion ($\log k_d \sim 0$), where rapid molecular mixing inhibits substantial Assembly growth.
        The inset shows $\Delta A$, highlighting a greater complexity difference between inflow and outflow in low-diffusion regimes compared to those with higher $k_d$.
        \textbf{(B)} Sensitivity for detecting complex molecules as a function of the diffusion rate ($k_d$) for the outflow reactor.
        The plot illustrates three distinct sensitivities, ranging from relative abundances of $10^{-3}$ (detecting one in $10^3$ molecules) to $10^{1}$ (one in 10), across a logarithmic range of $k_d$.
        Assembly indices of detectable molecules are shown in the inset.
    }
    \label{fig:total-assembly-and-thresholds}
\end{figure*}

Figure~\ref{fig:total-assembly-and-thresholds}A shows how both $A$ and $\Delta A$ evolve over time.
In the low diffusion regime ($\log k_d \sim -4$), total Assembly rises rapidly during the initial time steps and then gradually levels off as diffusion slowly redistributes molecules from the inflow to neighboring reactors.
Meanwhile, in the intermediate regime ($\log k_d \sim -2$), Assembly plateaus more quickly due to spatial dispersion limiting the frequency of constructive reactions.
Finally, in the high diffusion regime ($\log k_d \sim 0$), molecules spread almost immediately across the system, preventing a significant increase in Total Assembly.
The inset, showing $\Delta A$, supports this interpretation: the difference between inflow and outflow Assembly is much larger at low diffusion than in regimes with higher $k_d$ values.
In all cases the value of $\Delta A$ is positive, indicating $A$ is higher in the inflow than in the outflow.

The preceding results clearly demonstrate the impact of diffusion on the populations of chemical species and total Assembly $A$ in our model and on the distribution of chemical species across the different reactors.
As an increase in diffusion accelerates the distribution of molecules between reactors, it raises questions about the influence of the structure of the connections among them.


\paragraph{Sensitivities for detecting complex molecules.} 
A fundamental objective of this analysis is to help establish an upper limit on the expected assembly indices in the absence of biological systems.
Determining this upper boundary is crucial for life detection missions, as it enables the definition of specific thresholds aligned with the sensitivity parameters of detection instruments, such as mass spectrometers.

To investigate this, we conducted a series of simulations that replicate those used to generate the data in Figure~\ref{fig:varying-diffusion}.
In these new simulations, we assessed how the detection of high assembly molecules in the outflow reactor varies in response to changes in the diffusion rate.
Specifically, we sought to determine the upper limit, in terms of the integer $z$ and the assembly index, that can be detected within a single reactor. We varied the diffusion rate to examine how it might affect the detectable upper limit. We repeated the analysis using several detection sensitivities, thereby simulating coarser or finer resolutions representative of the instrumentation used for life-detection measurements on space missions.

Figure~\ref{fig:total-assembly-and-thresholds}B illustrates how logarithmic changes to the diffusion parameter affects these detections.
The analysis encompassed several distinct sensitivities, ranging from a relative abundance of $10^{-3}$ (i.e., the capacity to detect one in a thousand molecules) to $10^{-1}$ (the capacity to detect one in ten).
We evaluated the influence of diffusion on the detection of molecules in terms of AI (Figure~\ref{fig:total-assembly-and-thresholds}B, inset).

We now shift our focus to examining the impact of the reactor network's topology.
\color{black}


\paragraph{The Impact of Topology.}

\begin{figure}
    \centering
    \begin{overpic}[width=0.49\textwidth]{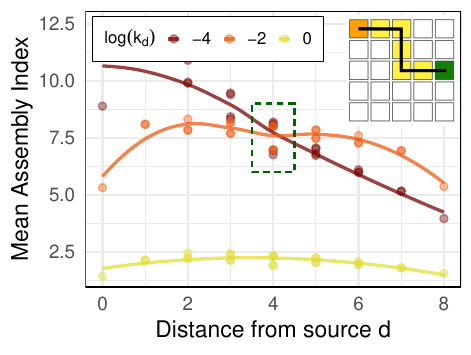}
    \end{overpic}
    \caption{
        Mean Assembly Index of molecules as a function of distance from the source reactor under three distinct diffusion regimes~—~low (red), medium (orange), and high (yellow).
        The green dashed box highlights an example of variability in the mean Assembly Index among reactors equidistant from the source ($d = 4$) under intermediate diffusion.
    }
    \label{fig:distance-from-source}
\end{figure}

To evaluate how the topology of a reactor network influences the dynamics of the underlying chemistry, we begin by comparing statistical moments of molecular distributions while varying network properties.
First, we have calculated the mean AI of molecules within individual chemostats, examining its relationship to the distance from the source reactor across three distinct diffusion regimes (Fig.~\ref{fig:distance-from-source}).

Our findings indicate that both diffusion and the exact location of reactors within the network significantly influence the mean AI.
Under low diffusion conditions, the mean AI peaks near the source reactor~—~where the majority of constructive reactions occur~—~and gradually declines with increasing distance from the source.
As diffusion intensifies, allowing molecules to disperse more broadly among chemostats, the impact of distance on the mean AI diminishes.
In scenarios characterized by high diffusion, the mean AI becomes nearly uniform across all reactors.

While reactor distance from the source correlates with the mean AI in that reactor, this relationship is not absolute.
Notably, reactors situated equidistant from the source consistently exhibit significantly different mean AI values.
For instance, at intermediate diffusion levels, multiple reactors located at a distance of $d=4$ from the source present divergent mean AI (Fig.~\ref{fig:distance-from-source}, green dashed box).
Consequently, distance from the source captures only a portion of the factors that control the distribution of AI in these systems.

[t]

To further investigate how topology influences the formation of high AI molecules, we focused on intermediate to high diffusion regimes.
These regimes are characterized by molecular populations that explore the entire spatial extent of the system.
We conducted an additional analysis to evaluate how overall topology affects the formation of complex species.
We generated two identical statistical ensembles of simulations; however, in one ensemble, we randomized the networks by pairwise edge swapping (see Section~\ref{subsec:randomization-procedure} in the Appendix for details).
This method preserved node degree and other low-level graph properties while eliminating high-level structural order, serving as a control.

\begin{figure*}[t]
	\centering
    \begin{overpic}[width=0.49\textwidth]{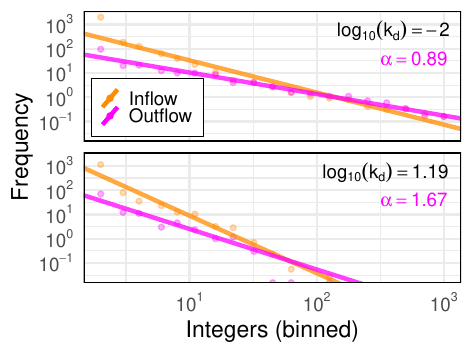}
        \put(0, 75){\textbf{(B)}}
    \end{overpic}
    \begin{overpic}[width=0.49\textwidth]{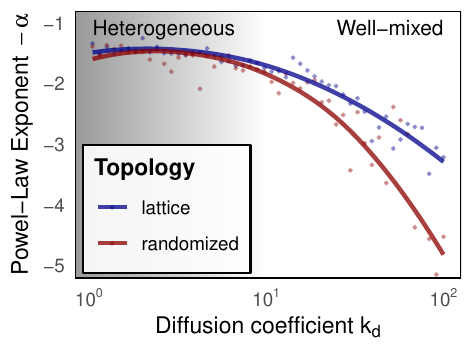}
        \put(0, 75){\textbf{(C)}}
    \end{overpic}

    \caption{
        \textbf{(B)} Examples of power-law fits for two representative simulations with $\log(k_d) = -2$ and $\log(k_d) = -1.19$.
        The orange curve represents the frequency of integers in the inflow, and the pink curve shows the outflow.
        Integers from 1 to 1{,}000 are grouped in bins of 50.
        Higher values of the exponent $\alpha$ correspond to steeper distributions with fewer complex species.
        \textbf{(C)} Power-law exponent $\alpha$ as a function of the diffusion parameter $k_d$ for both ordered lattice and randomized network topologies.
        A transition is observed from a heterogeneous regime at intermediate $k_d$ values to a well-mixed regime at high $k_d$ values.
        The more pronounced decrease in $\alpha$ for randomized topologies indicates a sharper decline in distribution tails, emphasizing the role of structural order in supporting the emergence of chemical complexity.
    }
    \label{fig:varying-topology}
\end{figure*}

To characterize the distribution tails, we binned the final populations up to $z = 1,000$ in bins of 50 integers and applied a power law fit of the form $ax^{-\alpha}$ (examples shown in Figure~\ref{fig:varying-topology}B).
Figure~\ref{fig:varying-topology}C illustrates how the power-law exponent varies as a function of the diffusion parameter $k_d$ for both the lattice and randomized configurations.

The results reveal a transition from a heterogeneous regime at intermediate $k_d$ values to a well-mixed regime at high $k_d$ values: as the system shifts to the homogeneous regime, the exponent $\alpha$ decreases more significantly in randomized topologies compared to lattice structures.
This trend in return signifies a faster decline in distribution tails, indicating fewer species with high AI.
In essence, our findings suggest that ordered topologies, such as regular lattices, facilitate the formation of relatively higher AI species, while randomized topologies suppresses them.





\section{Discussion} 
\label{sec:discussion}

Numerous physical environments are proposed to support the emergence of life, including hydrothermal vents, hot springs, and warm ponds.
These sites have been explored due to their potential to drive chemical complexity, a key factor in the origin of life.
It is believed that biomolecular precursors on primitive Earth interacted with diverse mineral substrates, potentially facilitating the emergence of life \citep{lambert_adsorption_2008}, numerical studies highlighting significant environmental effects on solution behaviors \citep{dujardin_formation_2023}.
For example, molecular dynamics simulations demonstrate that various mineral surfaces differently influence RNA pre-polymer formation.
Additionally, research shows that serpentinization~—~where specific rocks react with water, forming new minerals and releasing hydrogen~—~impacts confined fluids, thereby affecting abiotic organic production \citep{do_nascimento_vieira_ambivalent_2020, chogani_decoding_2023, lee_osmotic_2024, pital_analysis_2024}.
Porous media also play a role by influencing molecular diffusion \citep{tallarek_probing_2023, zhao_molecular_2024}, and diffusion processes can enhance both the diversity \citep{plum_spatial_2024} and functionality \citep{peng_two_2024} of autocatalytic networks. These constraints are important considerations for both the origin of life and for ensuring the reliability of life detection on other planets~\citep{foote_false_2023,jordan2024prebiotic}.

Our findings support the idea that several parameters associated with these systems significantly influence the distribution of molecular assembly indices.
Our simulations demonstrated that while an increase in diffusion could enhance the presence of certain simple molecules (e.g., dimers, Figure~\ref{fig:varying-diffusion}A-B), it nevertheless had a deterrent effect on constructive reactions.
This leads, on the one hand, to a decrease in the average molecular mass (Figure~\ref{fig:varying-diffusion}C) and, more generally, to a shift in the distribution of chemical species in our simulations (Figure~\ref{fig:varying-diffusion}D) and their AI (Figure~\ref{fig:varying-diffusion}D, inset).


Likewise, increasing the inflow significantly alters the dimer distribution by both boosting their numbers and shifting the diffusion coefficient required to achieve the highest dimer concentration (Figure~\ref{fig:varying-diffusion}B).
As more molecules are introduced, the likelihood of constructive (second-order) reactions rises quadratically.
The quadratic growth of constructive reactions, whose propensity depends on $n(n-1)$, surpasses the impact of diffusion reactions, which are first-order events and depend linearly on $n$.
This dynamic reduces the total molecule count (c.f.~Fig.~\ref{fig:varying-diffusion}C), necessitating higher diffusion coefficients to maintain a well-mixed regime that supports molecular complexity.
Put differently, this escalation leads to an overall increase in the average dimer count throughout the system, but overall AI distribution remains constrained by the diffusion regime the system is in.

Our analysis also made use of the notion of Assembly as defined in AT.
We employed this proposed measure to investigate variations in the chemical complexity at the inflow and outflow ends of the system (Figure~\ref{fig:total-assembly-and-thresholds}A).
We find that in all regimes that A reaches a finite steady-state, but our results reveal a certain paradox.
Shifting from a low-diffusion regime (Figure~\ref{fig:total-assembly-and-thresholds}A, red curve) to one where diffusion acts strongly (yellow curve) results in a trade-off between slower and more rapid exploration of the system’s topology, while maximum A is lower but reached earlier under high diffusion compared to the low-diffusion regime, where exploration is slower but leads to much higher AI chemical species.
This result may have very practical implications for the design of laboratory experiments on the origin of life.
For instance, it is possible that experiments with the highest short-term yield may not be those in which maximum complexity can be reached in the long term, indicating experiments that test for long term trends in molecule distributions may be more effective at identifying productive prebiotic chemstries ~\citep{matange2025evolution,asche_evidence_2024}.

We continued our analysis by calculating the maximum AI detectable within a single reactor for different sensitivities (Figure~\ref{fig:total-assembly-and-thresholds}B).
The observable trends result from two opposing effects: diffusion affects both the concentration of reactants~—~and thus their detectability~—~as well as the construction of molecules itself.
The detectability of high AI molecules therefore depends in a non-trivial way on these two transport-related effects within the system.
Our results suggest that analyzing samples at different resolutions may be meaningful: the detection of molecular complexity does not depend uniformly on diffusion across different proposed sensitivities.
For instance, a sensitivity of $\log\ [X] = -3$ (e.g. detection sensitivity of 1 in 1000 molecules) is much more sensitive to variations in the diffusion parameter than one of $\log\ [X] = -2$, which remains relatively stable in response to changes in $k_d$.
In the context of laboratory experiments, analyzing samples at different resolutions could therefore prove useful when diffusion is a significant constraint on the duration of the experiment.
\color{black}

Key transport parameters, such as the diffusion coefficient, which determine the extent to which molecules traverse the environmental topology, also amplify the topology’s impact on chemical processes.
We thus considered it essential to analyze the impact of the system’s topology, i.e., the structure of the connections between reactors, on the distribution of chemical species.
Our results show that the distance from the source reactor modulates the average AI of the molecules present in a reactor (Figure~\ref{fig:varying-topology}A).
This suggests that in porous materials, where catalytic surfaces are exposed to reactants to varying degrees, the resulting chemistry could more directly depend on transport parameters such as diffusion.
Moreover, the presence of fluctuations in the mean AI for reactors equidistant from the source reactor suggests that certain topological parameters could be optimized to produce the highest possible chemical complexity, with applications in origin-of-life research, bioengineering, and materials science.

The combined analysis of the impact of diffusion (Figure~\ref{fig:varying-diffusion}) and of a reactor’s distance from the source reactor (Figure~\ref{fig:varying-topology}A) further indicates that there exists an optimal regime for the construction of high AI molecules.
While both low-diffusion regimes (below $\log k_d \sim -4$) and high-diffusion regimes ($\log k_d \sim 0$ and beyond) limit the presence of high assembly index species throughout the system~—~either by confining the population to the inflow reactor or by accelerating their passage through the system~—~the intermediate diffusion regime ($\log k_d \sim -3$) promotes constructive reactions across the entire reactor network.

This exploration of the system’s topology by the population of chemical species (Figures~\ref{fig:varying-topology}B-C) showed that the structure of connections can also modulate the system’s dynamics.
We demonstrated that randomizing the system’s topology~—~thereby destroying any high-level properties of the network~—~constrains the construction of complex chemical species by shifting the molecular distribution toward simpler species.

This result aligns with several previous findings in complexity science suggesting that spatial patterning can facilitate the emergence of complexity.
We showed that the same process can apply in an artificial chemistry model: non-trivial topologies can modulate the formation of larger molecules.
The fact that the distribution of molecules generated by our simulations is well described by a power law further reflects the unavoidable trade-off between copy number and assembly indices: while our simulations do not yield an arbitrarily high assembly indices, they show that an abiotic process can nonetheless shift the maximum achievable threshold abiotic systems.
This, in turn, suggests that the development of life-detection methods based on assembly indices (e.g., \citep{marshall_identifying_2021}) should incorporate controls with non-trivial topological elements, in order to determine the boundaries of abiotic chemical systems.

We have previously highlighted that the regime optimizing the construction of complex molecules is that of intermediate diffusion, and this finding modulates the results obtained regarding the influence of topology on the formation of complex species.
In this intermediate regime, one might expect the influence of topology to be reduced if the occupancy state of reactors across the system is not saturated.
In other words, in the regime most favorable to broader AI distributions (that of intermediate diffusion), the positive influence of spatial patterning may not be as significant as in the high-diffusion regime, where the system efficiently explores the entire topology.
Furthermore, it is worth noting that the regime in which topology most strongly influences the upper threshold of assembly indices~—~the well-mixed, high-diffusion regime~—~is also the one in which the sensitivities examined earlier have the least impact on the maximum AI detected. In other words, topology has its greatest effect on the formation of higher AI in a regime where such molecules are relatively easier to detect even with lower sensitivity. This suggests that, in the context of space missions, for example, increasing instrumental sensitivity would not help distinguish false positives arising from the system’s topology. 
A possible implication of this would be that higher instrument resolution (e.g. the ability to distinguish molecules) may be more useful for confident life detection than instrument sensitivity (e.g. the ability to detect low abundance molecule), though this would require additional study.

Finally, the similarity of the exponent $\alpha$ for both lattice and randomized topologies in the heterogeneous regime, which determines the nature of the power law describing the distribution of integers, suggests that the increase in AI observed in this lower-diffusion regime is not directly driven by the type of topology connecting the reactors.
In this regime, the time spent in the system appears to be the determining factor in product distribution.
This motivates further investigation into the interplay between the diffusion regime and the influence of topology, insofar as these parameters may significantly modulate the emergence of high AI molecules.


\section{Conclusion}
\label{sec:conclusion}

The impact of spatial structure on the emergence of complexity (broadly understood) is a well-established result in complexity science.
Using an artificial chemistry model, we demonstrated that spatial structure can have profound implications for the behavior of a system of chemical reactors, and consequently for how we design experiments on the origin of life and its detection in the context of planetary missions.

Recent work on assembly theory provided empirical evidence suggesting that detectable quantities of sufficiently high AI molecules could serve as a method for life detection.
Our results suggest that experiments to determine the appropriate threshold for confident life detection may need to incorporate systems with non-trivial topologies, depending on the diffusion and transport parameters relevant to the target system.
Our results also show that several parameters characterizing environments of astrobiological interest~—~such as transport parameters and those related to the topology of porous materials~—~strongly influence the dynamics of chemical reactions and the resulting distributions of molecular species.
Our analysis suggests it is imperative to deepen our understanding of how these parameters interact with chemistry.
A long lineage of studies has examined the structure of networks in living systems; such analyses must now be extended to prebiotic chemistry to determine how various topologies~—~such as scale-free networks~—~might influence the emergence of molecular complexity.
Ultimately, this will contribute to a deeper comprehension of the environmental conditions necessary for the emergence of complexity, both on Earth and beyond.


\begin{acknowledgements}
	Computation of the model presented herein has been carried out on the Sol Supercomputer at Arizona State University \citep{jennewein_sol_2023}.
	
    We would like to thank Professor Sara Imari Walker, and Gage Siebert for insightful conversations at various stages of this project. 

    We would like to thank Professor Paul Charbonneau for his critical feedback and support on this project.

	This research was funded by the Natural Sciences and Engineering Research Council of Canada (Grant No. RGPIN/05278-2018), the Fonds de recherche Nature et Technologies of Québec (Grant No. 314488), the Fondation J. Armand Bombardier Excellence Scholarship, the Université de Montréal and the Center for Research in Astrophysics of Québec.

    C.M. acknowledges support from Human Frontiers in Science Program research grant RGEC29/2025.
\end{acknowledgements}

\paragraph{Author Contributions}
A.C.R. performed the analysis with support from C.M., developed the software with C.M., visualized results, drafted the initial manuscript, and edited and reviewed the final version. C.K. contributed to the project's conceptualization, preliminary analysis, and participated in final manuscript review and editing. C.M. conceptualized the study, supervised the project, developed the methodology and initial software prototypes, contributed to visualizations, and provided computational resources.

\appendix

\section{Supplementary Information}
\label{sec:SI}

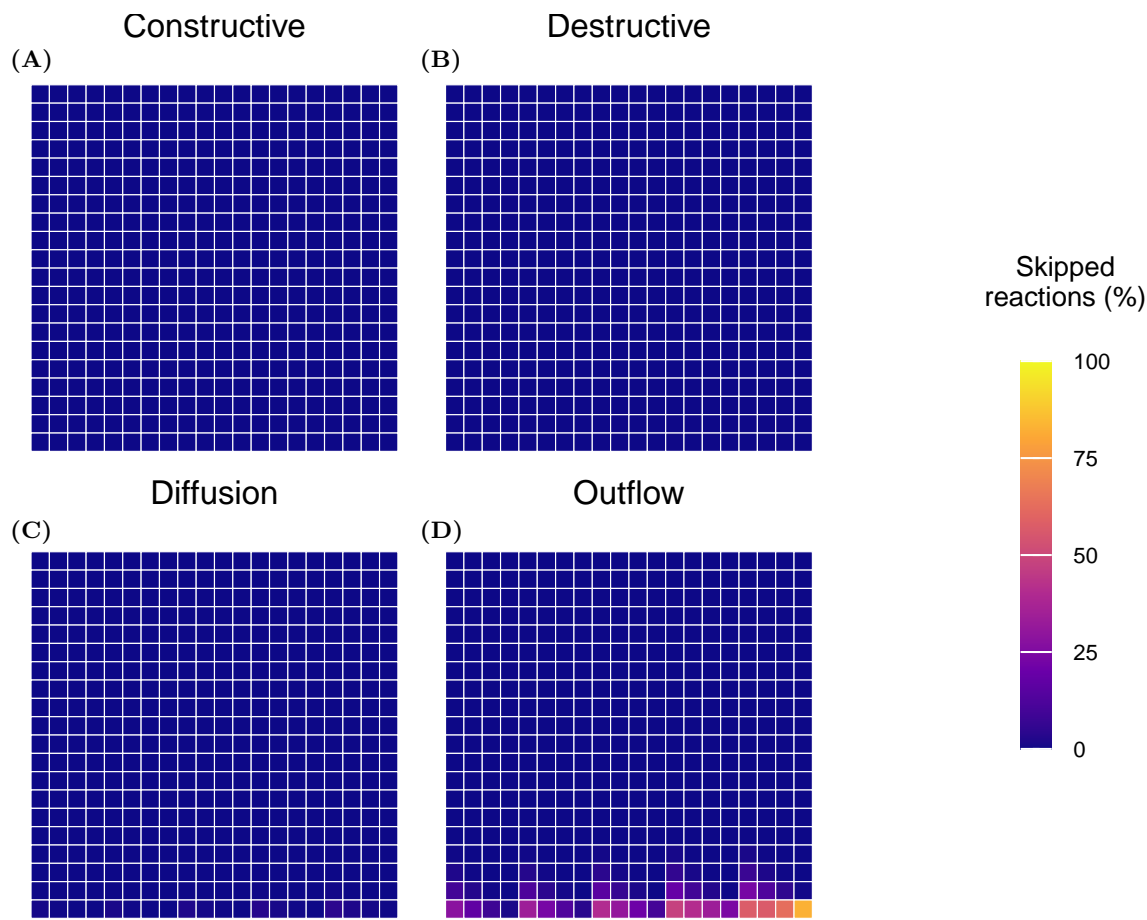
\begin{figure*}[t]
    \centering

    \begin{overpic}[width=1.00\textwidth]{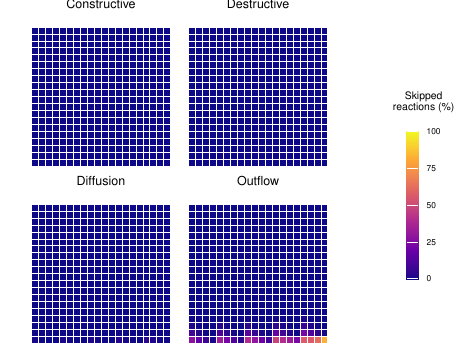}
        \put(5, 71){\textbf{(A)}}
        \put(38, 71){\textbf{(B)}}
        \put(5, 33){\textbf{(C)}}
        \put(38, 33){\textbf{(D)}}
    \end{overpic}

    \caption{
        Diagnostic plots showing the proportion of skipped reaction attempts for each reaction type:
        \textbf{(A)} Constructive (forward synthesis) reactions, 
        \textbf{(B)} Destructive (backward decomposition) reactions, 
        \textbf{(C)} Diffusion events, and 
        \textbf{(D)} Outflow events.
        Color intensity indicates the frequency of skipped reactions, which can result from low local reactant availability.
    }
    \label{fig:D01}
\end{figure*}

\subsection{Graph Randomization Procedure}
\label{subsec:randomization-procedure}

We used a directed edge-swapping algorithm to randomize the topology of a graph while preserving each node's in-degree and out-degree.
The procedure is as follows:


\begin{algorithm}[H] 
\caption{Edge‐Swapping Randomization}
\begin{algorithmic}[1]
  \For{$i = 1$ \textbf{to} $10\times E$}
    \State Randomly select edges $(u_1 \!\to\! v_1)$ and $(u_2 \!\to\! v_2)$ with all four nodes distinct
    \State Propose new edges $(u_1 \!\to\! v_2)$ and $(u_2 \!\to\! v_1)$
    \If{no self-loops \textbf{and} no duplicate edges}
      \State Replace the original edges with the proposed ones
    \EndIf
  \EndFor
\end{algorithmic}
\end{algorithm}

This process produces a randomized null model of the original network that preserves the full in-degree and out-degree sequence.
For each simulation, we applied this algorithm to the lattice topology, creating a new randomized version (i.e., 100 different randomized graphs for the 100 simulations of the ensemble).



\bibliography{flow-complexity.bib}

\begin{thebibliography}{44}%
\makeatletter
\providecommand \@ifxundefined [1]{%
 \@ifx{#1\undefined}
}%
\providecommand \@ifnum [1]{%
 \ifnum #1\expandafter \@firstoftwo
 \else \expandafter \@secondoftwo
 \fi
}%
\providecommand \@ifx [1]{%
 \ifx #1\expandafter \@firstoftwo
 \else \expandafter \@secondoftwo
 \fi
}%
\providecommand \natexlab [1]{#1}%
\providecommand \enquote  [1]{``#1''}%
\providecommand \bibnamefont  [1]{#1}%
\providecommand \bibfnamefont [1]{#1}%
\providecommand \citenamefont [1]{#1}%
\providecommand \href@noop [0]{\@secondoftwo}%
\providecommand \href [0]{\begingroup \@sanitize@url \@href}%
\providecommand \@href[1]{\@@startlink{#1}\@@href}%
\providecommand \@@href[1]{\endgroup#1\@@endlink}%
\providecommand \@sanitize@url [0]{\catcode `\\12\catcode `\$12\catcode
  `\&12\catcode `\#12\catcode `\^12\catcode `\_12\catcode `\%12\relax}%
\providecommand \@@startlink[1]{}%
\providecommand \@@endlink[0]{}%
\providecommand \url  [0]{\begingroup\@sanitize@url \@url }%
\providecommand \@url [1]{\endgroup\@href {#1}{\urlprefix }}%
\providecommand \urlprefix  [0]{URL }%
\providecommand \Eprint [0]{\href }%
\providecommand \doibase [0]{http://dx.doi.org/}%
\providecommand \selectlanguage [0]{\@gobble}%
\providecommand \bibinfo  [0]{\@secondoftwo}%
\providecommand \bibfield  [0]{\@secondoftwo}%
\providecommand \translation [1]{[#1]}%
\providecommand \BibitemOpen [0]{}%
\providecommand \bibitemStop [0]{}%
\providecommand \bibitemNoStop [0]{.\EOS\space}%
\providecommand \EOS [0]{\spacefactor3000\relax}%
\providecommand \BibitemShut  [1]{\csname bibitem#1\endcsname}%
\let\auto@bib@innerbib\@empty
\bibitem [{\citenamefont {Marshall}\ \emph {et~al.}(2021)\citenamefont
  {Marshall}, \citenamefont {Mathis}, \citenamefont {Carrick}, \citenamefont
  {Keenan}, \citenamefont {Cooper}, \citenamefont {Graham}, \citenamefont
  {Craven}, \citenamefont {Gromski}, \citenamefont {Moore}, \citenamefont
  {Walker},\ and\ \citenamefont {Cronin}}]{marshall_identifying_2021}%
  \BibitemOpen
  \bibfield  {author} {\bibinfo {author} {\bibfnamefont {S.~M.}\ \bibnamefont
  {Marshall}}, \bibinfo {author} {\bibfnamefont {C.}~\bibnamefont {Mathis}},
  \bibinfo {author} {\bibfnamefont {E.}~\bibnamefont {Carrick}}, \bibinfo
  {author} {\bibfnamefont {G.}~\bibnamefont {Keenan}}, \bibinfo {author}
  {\bibfnamefont {G.~J.~T.}\ \bibnamefont {Cooper}}, \bibinfo {author}
  {\bibfnamefont {H.}~\bibnamefont {Graham}}, \bibinfo {author} {\bibfnamefont
  {M.}~\bibnamefont {Craven}}, \bibinfo {author} {\bibfnamefont {P.~S.}\
  \bibnamefont {Gromski}}, \bibinfo {author} {\bibfnamefont {D.~G.}\
  \bibnamefont {Moore}}, \bibinfo {author} {\bibfnamefont {S.~I.}\ \bibnamefont
  {Walker}}, \ and\ \bibinfo {author} {\bibfnamefont {L.}~\bibnamefont
  {Cronin}},\ }\href {\doibase 10.1038/s41467-021-23258-x} {\bibfield
  {journal} {\bibinfo  {journal} {Nature Communications}\ }\textbf {\bibinfo
  {volume} {12}},\ \bibinfo {pages} {3033} (\bibinfo {year}
  {2021})}\BibitemShut {NoStop}%
\bibitem [{\citenamefont {Sharma}\ \emph {et~al.}(2023)\citenamefont {Sharma},
  \citenamefont {Cz{\'e}gel}, \citenamefont {Lachmann}, \citenamefont {Kempes},
  \citenamefont {Walker},\ and\ \citenamefont {Cronin}}]{sharma_assembly_2023}%
  \BibitemOpen
  \bibfield  {author} {\bibinfo {author} {\bibfnamefont {A.}~\bibnamefont
  {Sharma}}, \bibinfo {author} {\bibfnamefont {D.}~\bibnamefont {Cz{\'e}gel}},
  \bibinfo {author} {\bibfnamefont {M.}~\bibnamefont {Lachmann}}, \bibinfo
  {author} {\bibfnamefont {C.~P.}\ \bibnamefont {Kempes}}, \bibinfo {author}
  {\bibfnamefont {S.~I.}\ \bibnamefont {Walker}}, \ and\ \bibinfo {author}
  {\bibfnamefont {L.}~\bibnamefont {Cronin}},\ }\href {\doibase
  10.1038/s41586-023-06600-9} {\bibfield  {journal} {\bibinfo  {journal}
  {Nature}\ ,\ \bibinfo {pages} {1}} (\bibinfo {year} {2023})}\BibitemShut
  {NoStop}%
\bibitem [{\citenamefont {Jirasek}\ \emph {et~al.}(2023)\citenamefont
  {Jirasek}, \citenamefont {Sharma}, \citenamefont {Bame}, \citenamefont
  {Bell}, \citenamefont {Marshall}, \citenamefont {Mathis}, \citenamefont
  {Macleod}, \citenamefont {Cooper}, \citenamefont {Swart}, \citenamefont
  {Mollfulleda},\ and\ \citenamefont {Cronin}}]{jirasek_multimodal_2023}%
  \BibitemOpen
  \bibfield  {author} {\bibinfo {author} {\bibfnamefont {M.}~\bibnamefont
  {Jirasek}}, \bibinfo {author} {\bibfnamefont {A.}~\bibnamefont {Sharma}},
  \bibinfo {author} {\bibfnamefont {J.~R.}\ \bibnamefont {Bame}}, \bibinfo
  {author} {\bibfnamefont {N.}~\bibnamefont {Bell}}, \bibinfo {author}
  {\bibfnamefont {S.~M.}\ \bibnamefont {Marshall}}, \bibinfo {author}
  {\bibfnamefont {C.}~\bibnamefont {Mathis}}, \bibinfo {author} {\bibfnamefont
  {A.}~\bibnamefont {Macleod}}, \bibinfo {author} {\bibfnamefont {G.~J.~T.}\
  \bibnamefont {Cooper}}, \bibinfo {author} {\bibfnamefont {M.}~\bibnamefont
  {Swart}}, \bibinfo {author} {\bibfnamefont {R.}~\bibnamefont {Mollfulleda}},
  \ and\ \bibinfo {author} {\bibfnamefont {L.}~\bibnamefont {Cronin}},\ }\href
  {\doibase 10.48550/arXiv.2302.13753} {\enquote {\bibinfo {title} {Multimodal
  {{Techniques}} for {{Detecting Alien Life}} using {{Assembly Theory}} and
  {{Spectroscopy}}},}\ } (\bibinfo {year} {2023}),\ \Eprint
  {http://arxiv.org/abs/2302.13753} {arXiv:2302.13753 [physics, q-bio]}
  \BibitemShut {NoStop}%
\bibitem [{\citenamefont {Dragone}\ \emph {et~al.}(2021)\citenamefont
  {Dragone}, \citenamefont {Diaz}, \citenamefont {Hogg}, \citenamefont {Lyons},
  \citenamefont {Jackson}, \citenamefont {Wall}, \citenamefont {Adams},\ and\
  \citenamefont {Fierer}}]{dragone2021exploring}%
  \BibitemOpen
  \bibfield  {author} {\bibinfo {author} {\bibfnamefont {N.~B.}\ \bibnamefont
  {Dragone}}, \bibinfo {author} {\bibfnamefont {M.~A.}\ \bibnamefont {Diaz}},
  \bibinfo {author} {\bibfnamefont {I.~D.}\ \bibnamefont {Hogg}}, \bibinfo
  {author} {\bibfnamefont {W.~B.}\ \bibnamefont {Lyons}}, \bibinfo {author}
  {\bibfnamefont {W.~A.}\ \bibnamefont {Jackson}}, \bibinfo {author}
  {\bibfnamefont {D.~H.}\ \bibnamefont {Wall}}, \bibinfo {author}
  {\bibfnamefont {B.~J.}\ \bibnamefont {Adams}}, \ and\ \bibinfo {author}
  {\bibfnamefont {N.}~\bibnamefont {Fierer}},\ }\href@noop {} {\bibfield
  {journal} {\bibinfo  {journal} {Journal of Geophysical Research:
  Biogeosciences}\ }\textbf {\bibinfo {volume} {126}},\ \bibinfo {pages}
  {e2020JG006052} (\bibinfo {year} {2021})}\BibitemShut {NoStop}%
\bibitem [{\citenamefont {Ratliff}\ \emph {et~al.}(2023)\citenamefont
  {Ratliff}, \citenamefont {Fulford}, \citenamefont {Pozarycki}, \citenamefont
  {Wimp}, \citenamefont {Nichols}, \citenamefont {Osburn},\ and\ \citenamefont
  {Graham}}]{ratliff2023vacant}%
  \BibitemOpen
  \bibfield  {author} {\bibinfo {author} {\bibfnamefont {L.}~\bibnamefont
  {Ratliff}}, \bibinfo {author} {\bibfnamefont {A.}~\bibnamefont {Fulford}},
  \bibinfo {author} {\bibfnamefont {C.}~\bibnamefont {Pozarycki}}, \bibinfo
  {author} {\bibfnamefont {G.}~\bibnamefont {Wimp}}, \bibinfo {author}
  {\bibfnamefont {F.}~\bibnamefont {Nichols}}, \bibinfo {author} {\bibfnamefont
  {M.}~\bibnamefont {Osburn}}, \ and\ \bibinfo {author} {\bibfnamefont
  {H.}~\bibnamefont {Graham}},\ }\href@noop {} {\bibfield  {journal} {\bibinfo
  {journal} {bioRxiv}\ ,\ \bibinfo {pages} {2023}} (\bibinfo {year}
  {2023})}\BibitemShut {NoStop}%
\bibitem [{\citenamefont {Bell}\ \emph {et~al.}(2015)\citenamefont {Bell},
  \citenamefont {Boehnke}, \citenamefont {Harrison},\ and\ \citenamefont
  {Mao}}]{bell2015}%
  \BibitemOpen
  \bibfield  {author} {\bibinfo {author} {\bibfnamefont {E.~A.}\ \bibnamefont
  {Bell}}, \bibinfo {author} {\bibfnamefont {P.}~\bibnamefont {Boehnke}},
  \bibinfo {author} {\bibfnamefont {T.~M.}\ \bibnamefont {Harrison}}, \ and\
  \bibinfo {author} {\bibfnamefont {W.~L.}\ \bibnamefont {Mao}},\ }\href
  {\doibase 10.1073/pnas.1517557112} {\bibfield  {journal} {\bibinfo  {journal}
  {Proceedings of the National Academy of Sciences}\ }\textbf {\bibinfo
  {volume} {112}},\ \bibinfo {pages} {14518} (\bibinfo {year} {2015})},\
  \Eprint
  {http://arxiv.org/abs/https://www.pnas.org/doi/pdf/10.1073/pnas.1517557112}
  {https://www.pnas.org/doi/pdf/10.1073/pnas.1517557112} \BibitemShut {NoStop}%
\bibitem [{\citenamefont {Fortney}\ \emph {et~al.}(2019)\citenamefont
  {Fortney}, \citenamefont {Robinson}, \citenamefont {Domagal-Goldman},
  \citenamefont {Genio}, \citenamefont {Gordon}, \citenamefont {Gharib-Nezhad},
  \citenamefont {Lewis}, \citenamefont {Sousa-Silva}, \citenamefont
  {Airapetian}, \citenamefont {Drouin}, \citenamefont {Hargreaves},
  \citenamefont {Huang}, \citenamefont {Karman}, \citenamefont {Ramirez},
  \citenamefont {Rieker}, \citenamefont {Tennyson}, \citenamefont {Wordsworth},
  \citenamefont {Yurchenko}, \citenamefont {Johnson}, \citenamefont {Lee},
  \citenamefont {Dong}, \citenamefont {Kane}, \citenamefont {Lopez-Morales},
  \citenamefont {Fauchez}, \citenamefont {Lee}, \citenamefont {Marley},
  \citenamefont {Sung}, \citenamefont {Haghighipour}, \citenamefont {Robinson},
  \citenamefont {Horst}, \citenamefont {Gao}, \citenamefont {you Kao},
  \citenamefont {Dressing}, \citenamefont {Lupu}, \citenamefont {Savin},
  \citenamefont {Fleury}, \citenamefont {Venot}, \citenamefont {Ascenzi},
  \citenamefont {Milam}, \citenamefont {Linnartz}, \citenamefont {Gudipati},
  \citenamefont {Gronoff}, \citenamefont {Salama}, \citenamefont {Gavilan},
  \citenamefont {Bouwman}, \citenamefont {Turbet}, \citenamefont {Benilan},
  \citenamefont {Henderson}, \citenamefont {Batalha}, \citenamefont
  {Jensen-Clem}, \citenamefont {Lyons}, \citenamefont {Freedman}, \citenamefont
  {Schwieterman}, \citenamefont {Goyal}, \citenamefont {Mancini}, \citenamefont
  {Irwin}, \citenamefont {Desert}, \citenamefont {Molaverdikhani},
  \citenamefont {Gizis}, \citenamefont {Taylor}, \citenamefont {Lothringer},
  \citenamefont {Pierrehumbert}, \citenamefont {Zellem}, \citenamefont
  {Batalha}, \citenamefont {Rugheimer}, \citenamefont {Lustig-Yaeger},
  \citenamefont {Hu}, \citenamefont {Kempton}, \citenamefont {Arney},
  \citenamefont {Line}, \citenamefont {Alam}, \citenamefont {Moses},
  \citenamefont {Iro}, \citenamefont {Kreidberg}, \citenamefont {Blecic},
  \citenamefont {Louden}, \citenamefont {Molliere}, \citenamefont {Stevenson},
  \citenamefont {Swain}, \citenamefont {Bott}, \citenamefont {Madhusudhan},
  \citenamefont {Krissansen-Totton}, \citenamefont {Deming}, \citenamefont
  {Kitiashvili}, \citenamefont {Shkolnik}, \citenamefont {Rustamkulov},
  \citenamefont {Rogers},\ and\ \citenamefont
  {Close}}]{fortney2019needlaboratorymeasurementsab}%
  \BibitemOpen
  \bibfield  {author} {\bibinfo {author} {\bibfnamefont {J.~J.}\ \bibnamefont
  {Fortney}}, \bibinfo {author} {\bibfnamefont {T.~D.}\ \bibnamefont
  {Robinson}}, \bibinfo {author} {\bibfnamefont {S.}~\bibnamefont
  {Domagal-Goldman}}, \bibinfo {author} {\bibfnamefont {A.~D.~D.}\ \bibnamefont
  {Genio}}, \bibinfo {author} {\bibfnamefont {I.~E.}\ \bibnamefont {Gordon}},
  \bibinfo {author} {\bibfnamefont {E.}~\bibnamefont {Gharib-Nezhad}}, \bibinfo
  {author} {\bibfnamefont {N.}~\bibnamefont {Lewis}}, \bibinfo {author}
  {\bibfnamefont {C.}~\bibnamefont {Sousa-Silva}}, \bibinfo {author}
  {\bibfnamefont {V.}~\bibnamefont {Airapetian}}, \bibinfo {author}
  {\bibfnamefont {B.}~\bibnamefont {Drouin}}, \bibinfo {author} {\bibfnamefont
  {R.~J.}\ \bibnamefont {Hargreaves}}, \bibinfo {author} {\bibfnamefont
  {X.}~\bibnamefont {Huang}}, \bibinfo {author} {\bibfnamefont
  {T.}~\bibnamefont {Karman}}, \bibinfo {author} {\bibfnamefont {R.~M.}\
  \bibnamefont {Ramirez}}, \bibinfo {author} {\bibfnamefont {G.~B.}\
  \bibnamefont {Rieker}}, \bibinfo {author} {\bibfnamefont {J.}~\bibnamefont
  {Tennyson}}, \bibinfo {author} {\bibfnamefont {R.}~\bibnamefont
  {Wordsworth}}, \bibinfo {author} {\bibfnamefont {S.~N.}\ \bibnamefont
  {Yurchenko}}, \bibinfo {author} {\bibfnamefont {A.~V.}\ \bibnamefont
  {Johnson}}, \bibinfo {author} {\bibfnamefont {T.~J.}\ \bibnamefont {Lee}},
  \bibinfo {author} {\bibfnamefont {C.}~\bibnamefont {Dong}}, \bibinfo {author}
  {\bibfnamefont {S.}~\bibnamefont {Kane}}, \bibinfo {author} {\bibfnamefont
  {M.}~\bibnamefont {Lopez-Morales}}, \bibinfo {author} {\bibfnamefont
  {T.}~\bibnamefont {Fauchez}}, \bibinfo {author} {\bibfnamefont
  {T.}~\bibnamefont {Lee}}, \bibinfo {author} {\bibfnamefont {M.~S.}\
  \bibnamefont {Marley}}, \bibinfo {author} {\bibfnamefont {K.}~\bibnamefont
  {Sung}}, \bibinfo {author} {\bibfnamefont {N.}~\bibnamefont {Haghighipour}},
  \bibinfo {author} {\bibfnamefont {T.}~\bibnamefont {Robinson}}, \bibinfo
  {author} {\bibfnamefont {S.}~\bibnamefont {Horst}}, \bibinfo {author}
  {\bibfnamefont {P.}~\bibnamefont {Gao}}, \bibinfo {author} {\bibfnamefont
  {D.}~\bibnamefont {you Kao}}, \bibinfo {author} {\bibfnamefont
  {C.}~\bibnamefont {Dressing}}, \bibinfo {author} {\bibfnamefont
  {R.}~\bibnamefont {Lupu}}, \bibinfo {author} {\bibfnamefont {D.~W.}\
  \bibnamefont {Savin}}, \bibinfo {author} {\bibfnamefont {B.}~\bibnamefont
  {Fleury}}, \bibinfo {author} {\bibfnamefont {O.}~\bibnamefont {Venot}},
  \bibinfo {author} {\bibfnamefont {D.}~\bibnamefont {Ascenzi}}, \bibinfo
  {author} {\bibfnamefont {S.}~\bibnamefont {Milam}}, \bibinfo {author}
  {\bibfnamefont {H.}~\bibnamefont {Linnartz}}, \bibinfo {author}
  {\bibfnamefont {M.}~\bibnamefont {Gudipati}}, \bibinfo {author}
  {\bibfnamefont {G.}~\bibnamefont {Gronoff}}, \bibinfo {author} {\bibfnamefont
  {F.}~\bibnamefont {Salama}}, \bibinfo {author} {\bibfnamefont
  {L.}~\bibnamefont {Gavilan}}, \bibinfo {author} {\bibfnamefont
  {J.}~\bibnamefont {Bouwman}}, \bibinfo {author} {\bibfnamefont
  {M.}~\bibnamefont {Turbet}}, \bibinfo {author} {\bibfnamefont
  {Y.}~\bibnamefont {Benilan}}, \bibinfo {author} {\bibfnamefont
  {B.}~\bibnamefont {Henderson}}, \bibinfo {author} {\bibfnamefont
  {N.}~\bibnamefont {Batalha}}, \bibinfo {author} {\bibfnamefont
  {R.}~\bibnamefont {Jensen-Clem}}, \bibinfo {author} {\bibfnamefont
  {T.}~\bibnamefont {Lyons}}, \bibinfo {author} {\bibfnamefont
  {R.}~\bibnamefont {Freedman}}, \bibinfo {author} {\bibfnamefont
  {E.}~\bibnamefont {Schwieterman}}, \bibinfo {author} {\bibfnamefont
  {J.}~\bibnamefont {Goyal}}, \bibinfo {author} {\bibfnamefont
  {L.}~\bibnamefont {Mancini}}, \bibinfo {author} {\bibfnamefont
  {P.}~\bibnamefont {Irwin}}, \bibinfo {author} {\bibfnamefont {J.-M.}\
  \bibnamefont {Desert}}, \bibinfo {author} {\bibfnamefont {K.}~\bibnamefont
  {Molaverdikhani}}, \bibinfo {author} {\bibfnamefont {J.}~\bibnamefont
  {Gizis}}, \bibinfo {author} {\bibfnamefont {J.}~\bibnamefont {Taylor}},
  \bibinfo {author} {\bibfnamefont {J.}~\bibnamefont {Lothringer}}, \bibinfo
  {author} {\bibfnamefont {R.}~\bibnamefont {Pierrehumbert}}, \bibinfo {author}
  {\bibfnamefont {R.}~\bibnamefont {Zellem}}, \bibinfo {author} {\bibfnamefont
  {N.}~\bibnamefont {Batalha}}, \bibinfo {author} {\bibfnamefont
  {S.}~\bibnamefont {Rugheimer}}, \bibinfo {author} {\bibfnamefont
  {J.}~\bibnamefont {Lustig-Yaeger}}, \bibinfo {author} {\bibfnamefont
  {R.}~\bibnamefont {Hu}}, \bibinfo {author} {\bibfnamefont {E.}~\bibnamefont
  {Kempton}}, \bibinfo {author} {\bibfnamefont {G.}~\bibnamefont {Arney}},
  \bibinfo {author} {\bibfnamefont {M.}~\bibnamefont {Line}}, \bibinfo {author}
  {\bibfnamefont {M.}~\bibnamefont {Alam}}, \bibinfo {author} {\bibfnamefont
  {J.}~\bibnamefont {Moses}}, \bibinfo {author} {\bibfnamefont
  {N.}~\bibnamefont {Iro}}, \bibinfo {author} {\bibfnamefont {L.}~\bibnamefont
  {Kreidberg}}, \bibinfo {author} {\bibfnamefont {J.}~\bibnamefont {Blecic}},
  \bibinfo {author} {\bibfnamefont {T.}~\bibnamefont {Louden}}, \bibinfo
  {author} {\bibfnamefont {P.}~\bibnamefont {Molliere}}, \bibinfo {author}
  {\bibfnamefont {K.}~\bibnamefont {Stevenson}}, \bibinfo {author}
  {\bibfnamefont {M.}~\bibnamefont {Swain}}, \bibinfo {author} {\bibfnamefont
  {K.}~\bibnamefont {Bott}}, \bibinfo {author} {\bibfnamefont {N.}~\bibnamefont
  {Madhusudhan}}, \bibinfo {author} {\bibfnamefont {J.}~\bibnamefont
  {Krissansen-Totton}}, \bibinfo {author} {\bibfnamefont {D.}~\bibnamefont
  {Deming}}, \bibinfo {author} {\bibfnamefont {I.}~\bibnamefont {Kitiashvili}},
  \bibinfo {author} {\bibfnamefont {E.}~\bibnamefont {Shkolnik}}, \bibinfo
  {author} {\bibfnamefont {Z.}~\bibnamefont {Rustamkulov}}, \bibinfo {author}
  {\bibfnamefont {L.}~\bibnamefont {Rogers}}, \ and\ \bibinfo {author}
  {\bibfnamefont {L.}~\bibnamefont {Close}},\ }\href
  {https://arxiv.org/abs/1905.07064} {\enquote {\bibinfo {title} {The need for
  laboratory measurements and ab initio studies to aid understanding of
  exoplanetary atmospheres},}\ } (\bibinfo {year} {2019}),\ \Eprint
  {http://arxiv.org/abs/1905.07064} {arXiv:1905.07064 [astro-ph.EP]}
  \BibitemShut {NoStop}%
\bibitem [{\citenamefont {Fayolle}\ \emph {et~al.}(2020)\citenamefont
  {Fayolle}, \citenamefont {Barge}, \citenamefont {Cable}, \citenamefont
  {Drouin}, \citenamefont {Dworkin}, \citenamefont {Hanley}, \citenamefont
  {Henderson}, \citenamefont {Journaux}, \citenamefont {Noell}, \citenamefont
  {Salama}, \citenamefont {Sciamma-O'Brien}, \citenamefont {Waller},
  \citenamefont {Weber}, \citenamefont {Bennett}, \citenamefont {Blum},
  \citenamefont {Gudipati}, \citenamefont {Milam}, \citenamefont
  {Melwani-Daswani}, \citenamefont {Nuevo}, \citenamefont {Protopapa},\ and\
  \citenamefont {Smith}}]{fayolle2020criticallaboratorystudiesadvance}%
  \BibitemOpen
  \bibfield  {author} {\bibinfo {author} {\bibfnamefont {E.}~\bibnamefont
  {Fayolle}}, \bibinfo {author} {\bibfnamefont {L.}~\bibnamefont {Barge}},
  \bibinfo {author} {\bibfnamefont {M.}~\bibnamefont {Cable}}, \bibinfo
  {author} {\bibfnamefont {B.}~\bibnamefont {Drouin}}, \bibinfo {author}
  {\bibfnamefont {J.}~\bibnamefont {Dworkin}}, \bibinfo {author} {\bibfnamefont
  {J.}~\bibnamefont {Hanley}}, \bibinfo {author} {\bibfnamefont
  {B.}~\bibnamefont {Henderson}}, \bibinfo {author} {\bibfnamefont
  {B.}~\bibnamefont {Journaux}}, \bibinfo {author} {\bibfnamefont
  {A.}~\bibnamefont {Noell}}, \bibinfo {author} {\bibfnamefont
  {F.}~\bibnamefont {Salama}}, \bibinfo {author} {\bibfnamefont
  {E.}~\bibnamefont {Sciamma-O'Brien}}, \bibinfo {author} {\bibfnamefont
  {S.}~\bibnamefont {Waller}}, \bibinfo {author} {\bibfnamefont
  {J.}~\bibnamefont {Weber}}, \bibinfo {author} {\bibfnamefont
  {C.}~\bibnamefont {Bennett}}, \bibinfo {author} {\bibfnamefont
  {J.}~\bibnamefont {Blum}}, \bibinfo {author} {\bibfnamefont {M.}~\bibnamefont
  {Gudipati}}, \bibinfo {author} {\bibfnamefont {S.}~\bibnamefont {Milam}},
  \bibinfo {author} {\bibfnamefont {M.}~\bibnamefont {Melwani-Daswani}},
  \bibinfo {author} {\bibfnamefont {M.}~\bibnamefont {Nuevo}}, \bibinfo
  {author} {\bibfnamefont {S.}~\bibnamefont {Protopapa}}, \ and\ \bibinfo
  {author} {\bibfnamefont {R.}~\bibnamefont {Smith}},\ }\href
  {https://arxiv.org/abs/2009.12465} {\enquote {\bibinfo {title} {Critical
  laboratory studies to advance planetary science and support missions},}\ }
  (\bibinfo {year} {2020}),\ \Eprint {http://arxiv.org/abs/2009.12465}
  {arXiv:2009.12465 [astro-ph.IM]} \BibitemShut {NoStop}%
\bibitem [{\citenamefont {Matreux}\ \emph {et~al.}(2021)\citenamefont
  {Matreux}, \citenamefont {LeVay}, \citenamefont {Schmid}, \citenamefont
  {Aikkila}, \citenamefont {Belohlavek}, \citenamefont {{\c C}al{\i}{\c
  s}kano{\u g}lu}, \citenamefont {Salibi}, \citenamefont {K{\"u}hnlein},
  \citenamefont {Springsklee}, \citenamefont {Scheu}, \citenamefont {Dingwell},
  \citenamefont {Braun}, \citenamefont {Mutschler},\ and\ \citenamefont
  {Mast}}]{matreux_heat_2021}%
  \BibitemOpen
  \bibfield  {author} {\bibinfo {author} {\bibfnamefont {T.}~\bibnamefont
  {Matreux}}, \bibinfo {author} {\bibfnamefont {K.}~\bibnamefont {LeVay}},
  \bibinfo {author} {\bibfnamefont {A.}~\bibnamefont {Schmid}}, \bibinfo
  {author} {\bibfnamefont {P.}~\bibnamefont {Aikkila}}, \bibinfo {author}
  {\bibfnamefont {L.}~\bibnamefont {Belohlavek}}, \bibinfo {author}
  {\bibfnamefont {A.~Z.}\ \bibnamefont {{\c C}al{\i}{\c s}kano{\u g}lu}},
  \bibinfo {author} {\bibfnamefont {E.}~\bibnamefont {Salibi}}, \bibinfo
  {author} {\bibfnamefont {A.}~\bibnamefont {K{\"u}hnlein}}, \bibinfo {author}
  {\bibfnamefont {C.}~\bibnamefont {Springsklee}}, \bibinfo {author}
  {\bibfnamefont {B.}~\bibnamefont {Scheu}}, \bibinfo {author} {\bibfnamefont
  {D.~B.}\ \bibnamefont {Dingwell}}, \bibinfo {author} {\bibfnamefont
  {D.}~\bibnamefont {Braun}}, \bibinfo {author} {\bibfnamefont
  {H.}~\bibnamefont {Mutschler}}, \ and\ \bibinfo {author} {\bibfnamefont
  {C.~B.}\ \bibnamefont {Mast}},\ }\href {\doibase 10.1038/s41557-021-00772-5}
  {\bibfield  {journal} {\bibinfo  {journal} {Nature Chemistry}\ ,\ \bibinfo
  {pages} {1}} (\bibinfo {year} {2021})}\BibitemShut {NoStop}%
\bibitem [{\citenamefont {Matreux}\ \emph {et~al.}(2024)\citenamefont
  {Matreux}, \citenamefont {Aikkila}, \citenamefont {Scheu}, \citenamefont
  {Braun},\ and\ \citenamefont {Mast}}]{matreux_heat_2024}%
  \BibitemOpen
  \bibfield  {author} {\bibinfo {author} {\bibfnamefont {T.}~\bibnamefont
  {Matreux}}, \bibinfo {author} {\bibfnamefont {P.}~\bibnamefont {Aikkila}},
  \bibinfo {author} {\bibfnamefont {B.}~\bibnamefont {Scheu}}, \bibinfo
  {author} {\bibfnamefont {D.}~\bibnamefont {Braun}}, \ and\ \bibinfo {author}
  {\bibfnamefont {C.~B.}\ \bibnamefont {Mast}},\ }\href {\doibase
  10.1038/s41586-024-07193-7} {\bibfield  {journal} {\bibinfo  {journal}
  {Nature}\ }\textbf {\bibinfo {volume} {628}},\ \bibinfo {pages} {110}
  (\bibinfo {year} {2024})}\BibitemShut {NoStop}%
\bibitem [{\citenamefont {Turing}(1952)}]{turing_chemical_1952}%
  \BibitemOpen
  \bibfield  {author} {\bibinfo {author} {\bibfnamefont {A.~M.}\ \bibnamefont
  {Turing}},\ }\href {\doibase 10.1098/rstb.1952.0012} {\bibfield  {journal}
  {\bibinfo  {journal} {Philosophical Transactions of the Royal Society of
  London. Series B, Biological Sciences}\ }\textbf {\bibinfo {volume} {237}},\
  \bibinfo {pages} {37} (\bibinfo {year} {1952})},\ \Eprint
  {http://arxiv.org/abs/92463} {92463} \BibitemShut {NoStop}%
\bibitem [{\citenamefont {Boerlijst}\ and\ \citenamefont
  {Hogeweg}(1991)}]{boerlijst_spiral_1991}%
  \BibitemOpen
  \bibfield  {author} {\bibinfo {author} {\bibfnamefont {M.}~\bibnamefont
  {Boerlijst}}\ and\ \bibinfo {author} {\bibfnamefont {P.}~\bibnamefont
  {Hogeweg}},\ }\href {\doibase 10.1016/0167-2789(91)90049-F} {\bibfield
  {journal} {\bibinfo  {journal} {Physica D: Nonlinear Phenomena}\ }\textbf
  {\bibinfo {volume} {48}},\ \bibinfo {pages} {17} (\bibinfo {year}
  {1991})}\BibitemShut {NoStop}%
\bibitem [{\citenamefont {Colizzi}\ and\ \citenamefont
  {Hogeweg}(2016)}]{colizzi_parasites_2016}%
  \BibitemOpen
  \bibfield  {author} {\bibinfo {author} {\bibfnamefont {E.~S.}\ \bibnamefont
  {Colizzi}}\ and\ \bibinfo {author} {\bibfnamefont {P.}~\bibnamefont
  {Hogeweg}},\ }\href {\doibase 10.1371/journal.pcbi.1004902} {\bibfield
  {journal} {\bibinfo  {journal} {PLOS Computational Biology}\ }\textbf
  {\bibinfo {volume} {12}},\ \bibinfo {pages} {e1004902} (\bibinfo {year}
  {2016})}\BibitemShut {NoStop}%
\bibitem [{\citenamefont {Nowak}\ and\ \citenamefont
  {May}(1993)}]{nowak_spatial_1993}%
  \BibitemOpen
  \bibfield  {author} {\bibinfo {author} {\bibfnamefont {M.~A.}\ \bibnamefont
  {Nowak}}\ and\ \bibinfo {author} {\bibfnamefont {R.~M.}\ \bibnamefont
  {May}},\ }\href {\doibase 10.1142/S0218127493000040} {\bibfield  {journal}
  {\bibinfo  {journal} {International Journal of Bifurcation and Chaos}\
  }\textbf {\bibinfo {volume} {03}},\ \bibinfo {pages} {35} (\bibinfo {year}
  {1993})}\BibitemShut {NoStop}%
\bibitem [{\citenamefont {Lindgren}\ and\ \citenamefont
  {Nordahl}(1994)}]{lindgren_evolutionary_1994}%
  \BibitemOpen
  \bibfield  {author} {\bibinfo {author} {\bibfnamefont {K.}~\bibnamefont
  {Lindgren}}\ and\ \bibinfo {author} {\bibfnamefont {M.~G.}\ \bibnamefont
  {Nordahl}},\ }\href {\doibase 10.1016/0167-2789(94)90289-5} {\bibfield
  {journal} {\bibinfo  {journal} {Physica D: Nonlinear Phenomena}\ }\textbf
  {\bibinfo {volume} {75}},\ \bibinfo {pages} {292} (\bibinfo {year}
  {1994})}\BibitemShut {NoStop}%
\bibitem [{\citenamefont {Mizuuchi}\ and\ \citenamefont
  {Ichihashi}(2018)}]{mizuuchi_sustainable_2018}%
  \BibitemOpen
  \bibfield  {author} {\bibinfo {author} {\bibfnamefont {R.}~\bibnamefont
  {Mizuuchi}}\ and\ \bibinfo {author} {\bibfnamefont {N.}~\bibnamefont
  {Ichihashi}},\ }\href {\doibase 10.1038/s41559-018-0650-z} {\bibfield
  {journal} {\bibinfo  {journal} {Nature Ecology \& Evolution}\ }\textbf
  {\bibinfo {volume} {2}},\ \bibinfo {pages} {1654} (\bibinfo {year}
  {2018})}\BibitemShut {NoStop}%
\bibitem [{\citenamefont {{Champagne-Ruel}}\ and\ \citenamefont
  {Charbonneau}(2022)}]{champagne-ruel_mutation_2022}%
  \BibitemOpen
  \bibfield  {author} {\bibinfo {author} {\bibfnamefont {A.}~\bibnamefont
  {{Champagne-Ruel}}}\ and\ \bibinfo {author} {\bibfnamefont {P.}~\bibnamefont
  {Charbonneau}},\ }\href {\doibase 10.3390/life12020254} {\bibfield  {journal}
  {\bibinfo  {journal} {Life}\ }\textbf {\bibinfo {volume} {12}},\ \bibinfo
  {pages} {254} (\bibinfo {year} {2022})}\BibitemShut {NoStop}%
\bibitem [{\citenamefont {Asche}\ \emph {et~al.}(2024)\citenamefont {Asche},
  \citenamefont {Pow}, \citenamefont {Mehr}, \citenamefont {Cooper},
  \citenamefont {Sharma},\ and\ \citenamefont {Cronin}}]{asche_evidence_2024}%
  \BibitemOpen
  \bibfield  {author} {\bibinfo {author} {\bibfnamefont {S.}~\bibnamefont
  {Asche}}, \bibinfo {author} {\bibfnamefont {R.~W.}\ \bibnamefont {Pow}},
  \bibinfo {author} {\bibfnamefont {H.~M.}\ \bibnamefont {Mehr}}, \bibinfo
  {author} {\bibfnamefont {G.~J.~T.}\ \bibnamefont {Cooper}}, \bibinfo {author}
  {\bibfnamefont {A.}~\bibnamefont {Sharma}}, \ and\ \bibinfo {author}
  {\bibfnamefont {L.}~\bibnamefont {Cronin}},\ }\href {\doibase
  10.1002/syst.202400006} {\bibfield  {journal} {\bibinfo  {journal}
  {ChemSystemsChem}\ }\textbf {\bibinfo {volume} {n/a}},\ \bibinfo {pages}
  {e202400006} (\bibinfo {year} {2024})}\BibitemShut {NoStop}%
\bibitem [{\citenamefont {Hashiguchi}\ \emph {et~al.}(2023)\citenamefont
  {Hashiguchi}, \citenamefont {Aoki}, \citenamefont {Fukushima}, \citenamefont
  {Naraoka}, \citenamefont {Takano}, \citenamefont {Dworkin}, \citenamefont
  {Dworkin}, \citenamefont {Aponte}, \citenamefont {Elsila}, \citenamefont
  {Eiler}, \citenamefont {Furukawa}, \citenamefont {Furusho}, \citenamefont
  {Glavin}, \citenamefont {Graham}, \citenamefont {Hamase}, \citenamefont
  {Hertkorn}, \citenamefont {Isa}, \citenamefont {Koga}, \citenamefont
  {McLain}, \citenamefont {Mita}, \citenamefont {Oba}, \citenamefont {Ogawa},
  \citenamefont {Ohkouchi}, \citenamefont {{Orthous-Daunay}}, \citenamefont
  {Parker}, \citenamefont {Ruf}, \citenamefont {Sakai}, \citenamefont
  {{Schmitt-Kopplin}}, \citenamefont {Sugahara}, \citenamefont {Thissen},
  \citenamefont {Vuitton}, \citenamefont {Wolters}, \citenamefont {Yoshimura},
  \citenamefont {Yurimoto}, \citenamefont {Nakamura}, \citenamefont {Noguchi},
  \citenamefont {Okazaki}, \citenamefont {Yabuta}, \citenamefont {Sakamoto},
  \citenamefont {Tachibana}, \citenamefont {Yada}, \citenamefont {Nishimura},
  \citenamefont {Nakato}, \citenamefont {Miyazaki}, \citenamefont {Yogata},
  \citenamefont {Abe}, \citenamefont {Usui}, \citenamefont {Yoshikawa},
  \citenamefont {Saiki}, \citenamefont {Tanaka}, \citenamefont {Terui},
  \citenamefont {Nakazawa}, \citenamefont {Watanabe},\ and\ \citenamefont
  {Tsuda}}]{hashiguchi_spatial_2023}%
  \BibitemOpen
  \bibfield  {author} {\bibinfo {author} {\bibfnamefont {M.}~\bibnamefont
  {Hashiguchi}}, \bibinfo {author} {\bibfnamefont {D.}~\bibnamefont {Aoki}},
  \bibinfo {author} {\bibfnamefont {K.}~\bibnamefont {Fukushima}}, \bibinfo
  {author} {\bibfnamefont {H.}~\bibnamefont {Naraoka}}, \bibinfo {author}
  {\bibfnamefont {Y.}~\bibnamefont {Takano}}, \bibinfo {author} {\bibfnamefont
  {J.~P.}\ \bibnamefont {Dworkin}}, \bibinfo {author} {\bibfnamefont {K.~E.}\
  \bibnamefont {Dworkin}}, \bibinfo {author} {\bibfnamefont {J.~C.}\
  \bibnamefont {Aponte}}, \bibinfo {author} {\bibfnamefont {J.~E.}\
  \bibnamefont {Elsila}}, \bibinfo {author} {\bibfnamefont {J.~M.}\
  \bibnamefont {Eiler}}, \bibinfo {author} {\bibfnamefont {Y.}~\bibnamefont
  {Furukawa}}, \bibinfo {author} {\bibfnamefont {A.}~\bibnamefont {Furusho}},
  \bibinfo {author} {\bibfnamefont {D.~P.}\ \bibnamefont {Glavin}}, \bibinfo
  {author} {\bibfnamefont {H.~V.}\ \bibnamefont {Graham}}, \bibinfo {author}
  {\bibfnamefont {K.}~\bibnamefont {Hamase}}, \bibinfo {author} {\bibfnamefont
  {N.}~\bibnamefont {Hertkorn}}, \bibinfo {author} {\bibfnamefont
  {J.}~\bibnamefont {Isa}}, \bibinfo {author} {\bibfnamefont {T.}~\bibnamefont
  {Koga}}, \bibinfo {author} {\bibfnamefont {H.~L.}\ \bibnamefont {McLain}},
  \bibinfo {author} {\bibfnamefont {H.}~\bibnamefont {Mita}}, \bibinfo {author}
  {\bibfnamefont {Y.}~\bibnamefont {Oba}}, \bibinfo {author} {\bibfnamefont
  {N.~O.}\ \bibnamefont {Ogawa}}, \bibinfo {author} {\bibfnamefont
  {N.}~\bibnamefont {Ohkouchi}}, \bibinfo {author} {\bibfnamefont {F.-R.}\
  \bibnamefont {{Orthous-Daunay}}}, \bibinfo {author} {\bibfnamefont {E.~T.}\
  \bibnamefont {Parker}}, \bibinfo {author} {\bibfnamefont {A.}~\bibnamefont
  {Ruf}}, \bibinfo {author} {\bibfnamefont {S.}~\bibnamefont {Sakai}}, \bibinfo
  {author} {\bibfnamefont {P.}~\bibnamefont {{Schmitt-Kopplin}}}, \bibinfo
  {author} {\bibfnamefont {H.}~\bibnamefont {Sugahara}}, \bibinfo {author}
  {\bibfnamefont {R.}~\bibnamefont {Thissen}}, \bibinfo {author} {\bibfnamefont
  {V.}~\bibnamefont {Vuitton}}, \bibinfo {author} {\bibfnamefont
  {C.}~\bibnamefont {Wolters}}, \bibinfo {author} {\bibfnamefont
  {T.}~\bibnamefont {Yoshimura}}, \bibinfo {author} {\bibfnamefont
  {H.}~\bibnamefont {Yurimoto}}, \bibinfo {author} {\bibfnamefont
  {T.}~\bibnamefont {Nakamura}}, \bibinfo {author} {\bibfnamefont
  {T.}~\bibnamefont {Noguchi}}, \bibinfo {author} {\bibfnamefont
  {R.}~\bibnamefont {Okazaki}}, \bibinfo {author} {\bibfnamefont
  {H.}~\bibnamefont {Yabuta}}, \bibinfo {author} {\bibfnamefont
  {K.}~\bibnamefont {Sakamoto}}, \bibinfo {author} {\bibfnamefont
  {S.}~\bibnamefont {Tachibana}}, \bibinfo {author} {\bibfnamefont
  {T.}~\bibnamefont {Yada}}, \bibinfo {author} {\bibfnamefont {M.}~\bibnamefont
  {Nishimura}}, \bibinfo {author} {\bibfnamefont {A.}~\bibnamefont {Nakato}},
  \bibinfo {author} {\bibfnamefont {A.}~\bibnamefont {Miyazaki}}, \bibinfo
  {author} {\bibfnamefont {K.}~\bibnamefont {Yogata}}, \bibinfo {author}
  {\bibfnamefont {M.}~\bibnamefont {Abe}}, \bibinfo {author} {\bibfnamefont
  {T.}~\bibnamefont {Usui}}, \bibinfo {author} {\bibfnamefont {M.}~\bibnamefont
  {Yoshikawa}}, \bibinfo {author} {\bibfnamefont {T.}~\bibnamefont {Saiki}},
  \bibinfo {author} {\bibfnamefont {S.}~\bibnamefont {Tanaka}}, \bibinfo
  {author} {\bibfnamefont {F.}~\bibnamefont {Terui}}, \bibinfo {author}
  {\bibfnamefont {S.}~\bibnamefont {Nakazawa}}, \bibinfo {author}
  {\bibfnamefont {S.-i.}\ \bibnamefont {Watanabe}}, \ and\ \bibinfo {author}
  {\bibfnamefont {Y.}~\bibnamefont {Tsuda}},\ }\href {\doibase
  10.1186/s40623-023-01792-w} {\bibfield  {journal} {\bibinfo  {journal}
  {Earth, Planets and Space}\ }\textbf {\bibinfo {volume} {75}},\ \bibinfo
  {pages} {73} (\bibinfo {year} {2023})}\BibitemShut {NoStop}%
\bibitem [{\citenamefont {Rasmussen}\ \emph {et~al.}(2024)\citenamefont
  {Rasmussen}, \citenamefont {Muhling},\ and\ \citenamefont
  {Tosca}}]{rasmussen_nanoparticulate_2024}%
  \BibitemOpen
  \bibfield  {author} {\bibinfo {author} {\bibfnamefont {B.}~\bibnamefont
  {Rasmussen}}, \bibinfo {author} {\bibfnamefont {J.~R.}\ \bibnamefont
  {Muhling}}, \ and\ \bibinfo {author} {\bibfnamefont {N.~J.}\ \bibnamefont
  {Tosca}},\ }\href {\doibase 10.1126/sciadv.adj4789} {\bibfield  {journal}
  {\bibinfo  {journal} {Science Advances}\ }\textbf {\bibinfo {volume} {10}},\
  \bibinfo {pages} {eadj4789} (\bibinfo {year} {2024})}\BibitemShut {NoStop}%
\bibitem [{\citenamefont {Liu}\ and\ \citenamefont
  {Sumpter}(2018)}]{liu_mathematical_2018}%
  \BibitemOpen
  \bibfield  {author} {\bibinfo {author} {\bibfnamefont {Y.}~\bibnamefont
  {Liu}}\ and\ \bibinfo {author} {\bibfnamefont {D.~J.~T.}\ \bibnamefont
  {Sumpter}},\ }\href {\doibase 10.1074/jbc.RA118.003795} {\bibfield  {journal}
  {\bibinfo  {journal} {Journal of Biological Chemistry}\ }\textbf {\bibinfo
  {volume} {293}},\ \bibinfo {pages} {18854} (\bibinfo {year}
  {2018})}\BibitemShut {NoStop}%
\bibitem [{\citenamefont {Banzhaf}\ and\ \citenamefont
  {Yamamoto}(2015)}]{banzhaf_artificial_2015}%
  \BibitemOpen
  \bibfield  {author} {\bibinfo {author} {\bibfnamefont {W.}~\bibnamefont
  {Banzhaf}}\ and\ \bibinfo {author} {\bibfnamefont {L.}~\bibnamefont
  {Yamamoto}},\ }\href@noop {} {\emph {\bibinfo {title} {Artificial
  Chemistries}}}\ (\bibinfo  {publisher} {The MIT Press},\ \bibinfo {address}
  {Cambridge, MA},\ \bibinfo {year} {2015})\BibitemShut {NoStop}%
\bibitem [{\citenamefont {Albert}\ and\ \citenamefont
  {Barab{\'a}si}(2002)}]{albert_statistical_2002}%
  \BibitemOpen
  \bibfield  {author} {\bibinfo {author} {\bibfnamefont {R.}~\bibnamefont
  {Albert}}\ and\ \bibinfo {author} {\bibfnamefont {A.-L.}\ \bibnamefont
  {Barab{\'a}si}},\ }\href {\doibase 10.1103/RevModPhys.74.47} {\bibfield
  {journal} {\bibinfo  {journal} {Reviews of modern physics}\ }\textbf
  {\bibinfo {volume} {74}},\ \bibinfo {pages} {47} (\bibinfo {year}
  {2002})}\BibitemShut {NoStop}%
\bibitem [{\citenamefont {Barab{\'a}si}\ and\ \citenamefont
  {P{\'o}sfai}(2016)}]{barabasi_network_2016}%
  \BibitemOpen
  \bibfield  {author} {\bibinfo {author} {\bibfnamefont {A.-L.}\ \bibnamefont
  {Barab{\'a}si}}\ and\ \bibinfo {author} {\bibfnamefont {M.}~\bibnamefont
  {P{\'o}sfai}},\ }\href@noop {} {\emph {\bibinfo {title} {Network
  {{Science}}}}},\ \bibinfo {edition} {1st}\ ed.\ (\bibinfo  {publisher}
  {Cambridge University Press},\ \bibinfo {address} {Cambridge, United
  Kingdom},\ \bibinfo {year} {2016})\BibitemShut {NoStop}%
\bibitem [{\citenamefont {{Rodriguez-Garcia}}\ \emph
  {et~al.}(2015)\citenamefont {{Rodriguez-Garcia}}, \citenamefont {Surman},
  \citenamefont {Cooper}, \citenamefont {{Su{\'a}rez-Marina}}, \citenamefont
  {Hosni}, \citenamefont {Lee},\ and\ \citenamefont
  {Cronin}}]{rodriguez-garcia_formation_2015}%
  \BibitemOpen
  \bibfield  {author} {\bibinfo {author} {\bibfnamefont {M.}~\bibnamefont
  {{Rodriguez-Garcia}}}, \bibinfo {author} {\bibfnamefont {A.~J.}\ \bibnamefont
  {Surman}}, \bibinfo {author} {\bibfnamefont {G.~J.~T.}\ \bibnamefont
  {Cooper}}, \bibinfo {author} {\bibfnamefont {I.}~\bibnamefont
  {{Su{\'a}rez-Marina}}}, \bibinfo {author} {\bibfnamefont {Z.}~\bibnamefont
  {Hosni}}, \bibinfo {author} {\bibfnamefont {M.~P.}\ \bibnamefont {Lee}}, \
  and\ \bibinfo {author} {\bibfnamefont {L.}~\bibnamefont {Cronin}},\ }\href
  {\doibase 10.1038/ncomms9385} {\bibfield  {journal} {\bibinfo  {journal}
  {Nature Communications}\ }\textbf {\bibinfo {volume} {6}},\ \bibinfo {pages}
  {8385} (\bibinfo {year} {2015})}\BibitemShut {NoStop}%
\bibitem [{\citenamefont {Gillespie}(1977)}]{gillespie_exact_1977}%
  \BibitemOpen
  \bibfield  {author} {\bibinfo {author} {\bibfnamefont {D.~T.}\ \bibnamefont
  {Gillespie}},\ }\href {\doibase 10.1021/j100540a008} {\bibfield  {journal}
  {\bibinfo  {journal} {The Journal of Physical Chemistry}\ }\textbf {\bibinfo
  {volume} {81}},\ \bibinfo {pages} {2340} (\bibinfo {year}
  {1977})}\BibitemShut {NoStop}%
\bibitem [{\citenamefont {Gillespie}(2001)}]{gillespie_approximate_2001}%
  \BibitemOpen
  \bibfield  {author} {\bibinfo {author} {\bibfnamefont {D.~T.}\ \bibnamefont
  {Gillespie}},\ }\href {\doibase 10.1063/1.1378322} {\bibfield  {journal}
  {\bibinfo  {journal} {The Journal of Chemical Physics}\ }\textbf {\bibinfo
  {volume} {115}},\ \bibinfo {pages} {1716} (\bibinfo {year}
  {2001})}\BibitemShut {NoStop}%
\bibitem [{\citenamefont {Marshall}\ \emph {et~al.}(2022)\citenamefont
  {Marshall}, \citenamefont {Moore}, \citenamefont {Murray}, \citenamefont
  {Walker},\ and\ \citenamefont {Cronin}}]{marshall_formalising_2022}%
  \BibitemOpen
  \bibfield  {author} {\bibinfo {author} {\bibfnamefont {S.~M.}\ \bibnamefont
  {Marshall}}, \bibinfo {author} {\bibfnamefont {D.~G.}\ \bibnamefont {Moore}},
  \bibinfo {author} {\bibfnamefont {A.~R.~G.}\ \bibnamefont {Murray}}, \bibinfo
  {author} {\bibfnamefont {S.~I.}\ \bibnamefont {Walker}}, \ and\ \bibinfo
  {author} {\bibfnamefont {L.}~\bibnamefont {Cronin}},\ }\href {\doibase
  10.3390/e24070884} {\bibfield  {journal} {\bibinfo  {journal} {Entropy}\
  }\textbf {\bibinfo {volume} {24}},\ \bibinfo {pages} {884} (\bibinfo {year}
  {2022})}\BibitemShut {NoStop}%
\bibitem [{\citenamefont {Thurber}(1999)}]{thurber_efficient_1999}%
  \BibitemOpen
  \bibfield  {author} {\bibinfo {author} {\bibfnamefont {E.~G.}\ \bibnamefont
  {Thurber}},\ }\href {\doibase 10.1137/S0097539795295663} {\bibfield
  {journal} {\bibinfo  {journal} {SIAM Journal on Computing}\ }\textbf
  {\bibinfo {volume} {28}},\ \bibinfo {pages} {1247} (\bibinfo {year}
  {1999})}\BibitemShut {NoStop}%
\bibitem [{\citenamefont {Toffoli}\ and\ \citenamefont
  {Margolus}(1987)}]{toffoli1987cellular}%
  \BibitemOpen
  \bibfield  {author} {\bibinfo {author} {\bibfnamefont {T.}~\bibnamefont
  {Toffoli}}\ and\ \bibinfo {author} {\bibfnamefont {N.}~\bibnamefont
  {Margolus}},\ }\href@noop {} {\emph {\bibinfo {title} {Cellular automata
  machines: a new environment for modeling}}}\ (\bibinfo  {publisher} {MIT
  press},\ \bibinfo {year} {1987})\BibitemShut {NoStop}%
\bibitem [{\citenamefont {Lambert}(2008)}]{lambert_adsorption_2008}%
  \BibitemOpen
  \bibfield  {author} {\bibinfo {author} {\bibfnamefont {J.-F.}\ \bibnamefont
  {Lambert}},\ }\href {\doibase 10.1007/s11084-008-9128-3} {\bibfield
  {journal} {\bibinfo  {journal} {Origins of Life and Evolution of Biospheres}\
  }\textbf {\bibinfo {volume} {38}},\ \bibinfo {pages} {211} (\bibinfo {year}
  {2008})}\BibitemShut {NoStop}%
\bibitem [{\citenamefont {Dujardin}\ \emph {et~al.}(2023)\citenamefont
  {Dujardin}, \citenamefont {Himbert}, \citenamefont {Pudritz},\ and\
  \citenamefont {Rheinst{\"a}dter}}]{dujardin_formation_2023}%
  \BibitemOpen
  \bibfield  {author} {\bibinfo {author} {\bibfnamefont {A.}~\bibnamefont
  {Dujardin}}, \bibinfo {author} {\bibfnamefont {S.}~\bibnamefont {Himbert}},
  \bibinfo {author} {\bibfnamefont {R.}~\bibnamefont {Pudritz}}, \ and\
  \bibinfo {author} {\bibfnamefont {M.~C.}\ \bibnamefont {Rheinst{\"a}dter}},\
  }\href {\doibase 10.3390/life13010112} {\bibfield  {journal} {\bibinfo
  {journal} {Life}\ }\textbf {\bibinfo {volume} {13}},\ \bibinfo {pages} {112}
  (\bibinfo {year} {2023})}\BibitemShut {NoStop}%
\bibitem [{\citenamefont {{do Nascimento Vieira}}\ \emph
  {et~al.}(2020)\citenamefont {{do Nascimento Vieira}}, \citenamefont
  {Kleinermanns}, \citenamefont {Martin},\ and\ \citenamefont
  {Preiner}}]{do_nascimento_vieira_ambivalent_2020}%
  \BibitemOpen
  \bibfield  {author} {\bibinfo {author} {\bibfnamefont {A.}~\bibnamefont {{do
  Nascimento Vieira}}}, \bibinfo {author} {\bibfnamefont {K.}~\bibnamefont
  {Kleinermanns}}, \bibinfo {author} {\bibfnamefont {W.~F.}\ \bibnamefont
  {Martin}}, \ and\ \bibinfo {author} {\bibfnamefont {M.}~\bibnamefont
  {Preiner}},\ }\href {\doibase 10.1002/1873-3468.13815} {\bibfield  {journal}
  {\bibinfo  {journal} {FEBS Letters}\ }\textbf {\bibinfo {volume} {594}},\
  \bibinfo {pages} {2717} (\bibinfo {year} {2020})}\BibitemShut {NoStop}%
\bibitem [{\citenamefont {Chogani}\ and\ \citenamefont
  {Pl{\"u}mper}(2023)}]{chogani_decoding_2023}%
  \BibitemOpen
  \bibfield  {author} {\bibinfo {author} {\bibfnamefont {A.}~\bibnamefont
  {Chogani}}\ and\ \bibinfo {author} {\bibfnamefont {O.}~\bibnamefont
  {Pl{\"u}mper}},\ }\href {\doibase 10.1007/s00410-023-02062-4} {\bibfield
  {journal} {\bibinfo  {journal} {Contributions to Mineralogy and Petrology}\
  }\textbf {\bibinfo {volume} {178}},\ \bibinfo {pages} {78} (\bibinfo {year}
  {2023})}\BibitemShut {NoStop}%
\bibitem [{\citenamefont {Lee}\ \emph {et~al.}(2024)\citenamefont {Lee},
  \citenamefont {Okumura}, \citenamefont {Ooka}, \citenamefont {Adachi},
  \citenamefont {Hikima}, \citenamefont {Hirata}, \citenamefont {Kawano},
  \citenamefont {Matsuura}, \citenamefont {Yamamoto}, \citenamefont {Yamamoto},
  \citenamefont {Yamaguchi}, \citenamefont {Lee}, \citenamefont {Takahashi},
  \citenamefont {Nam}, \citenamefont {Ohara}, \citenamefont {Hashizume},
  \citenamefont {McGlynn},\ and\ \citenamefont {Nakamura}}]{lee_osmotic_2024}%
  \BibitemOpen
  \bibfield  {author} {\bibinfo {author} {\bibfnamefont {H.-E.}\ \bibnamefont
  {Lee}}, \bibinfo {author} {\bibfnamefont {T.}~\bibnamefont {Okumura}},
  \bibinfo {author} {\bibfnamefont {H.}~\bibnamefont {Ooka}}, \bibinfo {author}
  {\bibfnamefont {K.}~\bibnamefont {Adachi}}, \bibinfo {author} {\bibfnamefont
  {T.}~\bibnamefont {Hikima}}, \bibinfo {author} {\bibfnamefont
  {K.}~\bibnamefont {Hirata}}, \bibinfo {author} {\bibfnamefont
  {Y.}~\bibnamefont {Kawano}}, \bibinfo {author} {\bibfnamefont
  {H.}~\bibnamefont {Matsuura}}, \bibinfo {author} {\bibfnamefont
  {M.}~\bibnamefont {Yamamoto}}, \bibinfo {author} {\bibfnamefont
  {M.}~\bibnamefont {Yamamoto}}, \bibinfo {author} {\bibfnamefont
  {A.}~\bibnamefont {Yamaguchi}}, \bibinfo {author} {\bibfnamefont {J.-E.}\
  \bibnamefont {Lee}}, \bibinfo {author} {\bibfnamefont {H.}~\bibnamefont
  {Takahashi}}, \bibinfo {author} {\bibfnamefont {K.~T.}\ \bibnamefont {Nam}},
  \bibinfo {author} {\bibfnamefont {Y.}~\bibnamefont {Ohara}}, \bibinfo
  {author} {\bibfnamefont {D.}~\bibnamefont {Hashizume}}, \bibinfo {author}
  {\bibfnamefont {S.~E.}\ \bibnamefont {McGlynn}}, \ and\ \bibinfo {author}
  {\bibfnamefont {R.}~\bibnamefont {Nakamura}},\ }\href {\doibase
  10.1038/s41467-024-52332-3} {\bibfield  {journal} {\bibinfo  {journal}
  {Nature Communications}\ }\textbf {\bibinfo {volume} {15}},\ \bibinfo {pages}
  {8193} (\bibinfo {year} {2024})}\BibitemShut {NoStop}%
\bibitem [{\citenamefont {Pital}\ \emph {et~al.}(2024)\citenamefont {Pital},
  \citenamefont {Bromley}, \citenamefont {Dorn}, \citenamefont {Kim},\ and\
  \citenamefont {Stockton}}]{pital_analysis_2024}%
  \BibitemOpen
  \bibfield  {author} {\bibinfo {author} {\bibfnamefont {A.}~\bibnamefont
  {Pital}}, \bibinfo {author} {\bibfnamefont {M.}~\bibnamefont {Bromley}},
  \bibinfo {author} {\bibfnamefont {M.}~\bibnamefont {Dorn}}, \bibinfo {author}
  {\bibfnamefont {J.}~\bibnamefont {Kim}}, \ and\ \bibinfo {author}
  {\bibfnamefont {A.}~\bibnamefont {Stockton}},\ }\href {\doibase
  10.1089/ast.2021.0088} {\bibfield  {journal} {\bibinfo  {journal}
  {Astrobiology}\ }\textbf {\bibinfo {volume} {24}},\ \bibinfo {pages} {138}
  (\bibinfo {year} {2024})}\BibitemShut {NoStop}%
\bibitem [{\citenamefont {Tallarek}\ \emph {et~al.}(2023)\citenamefont
  {Tallarek}, \citenamefont {Hlushkou}, \citenamefont {Trebel},\ and\
  \citenamefont {H{\"o}ltzel}}]{tallarek_probing_2023}%
  \BibitemOpen
  \bibfield  {author} {\bibinfo {author} {\bibfnamefont {U.}~\bibnamefont
  {Tallarek}}, \bibinfo {author} {\bibfnamefont {D.}~\bibnamefont {Hlushkou}},
  \bibinfo {author} {\bibfnamefont {N.}~\bibnamefont {Trebel}}, \ and\ \bibinfo
  {author} {\bibfnamefont {A.}~\bibnamefont {H{\"o}ltzel}},\ }\href {\doibase
  10.1002/cite.202300027} {\bibfield  {journal} {\bibinfo  {journal} {Chemie
  Ingenieur Technik}\ }\textbf {\bibinfo {volume} {95}},\ \bibinfo {pages}
  {1777} (\bibinfo {year} {2023})}\BibitemShut {NoStop}%
\bibitem [{\citenamefont {Zhao}\ \emph {et~al.}(2024)\citenamefont {Zhao},
  \citenamefont {Xue}, \citenamefont {Lu}, \citenamefont {Greenwell},
  \citenamefont {Xu}, \citenamefont {He},\ and\ \citenamefont
  {Erastova}}]{zhao_molecular_2024}%
  \BibitemOpen
  \bibfield  {author} {\bibinfo {author} {\bibfnamefont {R.}~\bibnamefont
  {Zhao}}, \bibinfo {author} {\bibfnamefont {H.}~\bibnamefont {Xue}}, \bibinfo
  {author} {\bibfnamefont {S.}~\bibnamefont {Lu}}, \bibinfo {author}
  {\bibfnamefont {H.~C.}\ \bibnamefont {Greenwell}}, \bibinfo {author}
  {\bibfnamefont {Y.}~\bibnamefont {Xu}}, \bibinfo {author} {\bibfnamefont
  {T.}~\bibnamefont {He}}, \ and\ \bibinfo {author} {\bibfnamefont
  {V.}~\bibnamefont {Erastova}},\ }\href {\doibase 10.1063/5.0226864}
  {\bibfield  {journal} {\bibinfo  {journal} {Physics of Fluids}\ }\textbf
  {\bibinfo {volume} {36}},\ \bibinfo {pages} {092027} (\bibinfo {year}
  {2024})}\BibitemShut {NoStop}%
\bibitem [{\citenamefont {Plum}\ \emph {et~al.}(2024)\citenamefont {Plum},
  \citenamefont {Kempes}, \citenamefont {Peng},\ and\ \citenamefont
  {Baum}}]{plum_spatial_2024}%
  \BibitemOpen
  \bibfield  {author} {\bibinfo {author} {\bibfnamefont {A.~M.}\ \bibnamefont
  {Plum}}, \bibinfo {author} {\bibfnamefont {C.~P.}\ \bibnamefont {Kempes}},
  \bibinfo {author} {\bibfnamefont {Z.}~\bibnamefont {Peng}}, \ and\ \bibinfo
  {author} {\bibfnamefont {D.~A.}\ \bibnamefont {Baum}},\ }\href {\doibase
  10.48550/arXiv.2212.14445} {\enquote {\bibinfo {title} {Spatial {{Structure
  Supports Diversity}} in {{Prebiotic Autocatalytic Chemical Ecosystems}}},}\ }
  (\bibinfo {year} {2024}),\ \Eprint {http://arxiv.org/abs/2212.14445}
  {arXiv:2212.14445 [q-bio]} \BibitemShut {NoStop}%
\bibitem [{\citenamefont {Peng}\ and\ \citenamefont
  {Adam}(2024)}]{peng_two_2024}%
  \BibitemOpen
  \bibfield  {author} {\bibinfo {author} {\bibfnamefont {Z.}~\bibnamefont
  {Peng}}\ and\ \bibinfo {author} {\bibfnamefont {Z.~R.}\ \bibnamefont
  {Adam}},\ }\href {\doibase 10.1016/j.chaos.2024.114955} {\bibfield  {journal}
  {\bibinfo  {journal} {Chaos, Solitons \& Fractals}\ }\textbf {\bibinfo
  {volume} {184}},\ \bibinfo {pages} {114955} (\bibinfo {year}
  {2024})}\BibitemShut {NoStop}%
\bibitem [{\citenamefont {Foote}\ \emph {et~al.}(2023)\citenamefont {Foote},
  \citenamefont {Sinhadc}, \citenamefont {Mathis},\ and\ \citenamefont
  {Walker}}]{foote_false_2023}%
  \BibitemOpen
  \bibfield  {author} {\bibinfo {author} {\bibfnamefont {S.}~\bibnamefont
  {Foote}}, \bibinfo {author} {\bibfnamefont {P.}~\bibnamefont {Sinhadc}},
  \bibinfo {author} {\bibfnamefont {C.}~\bibnamefont {Mathis}}, \ and\ \bibinfo
  {author} {\bibfnamefont {S.~I.}\ \bibnamefont {Walker}},\ }\href {\doibase
  10.1089/ast.2023.0005} {\bibfield  {journal} {\bibinfo  {journal}
  {Astrobiology}\ }\textbf {\bibinfo {volume} {23}},\ \bibinfo {pages} {1189}
  (\bibinfo {year} {2023})}\BibitemShut {NoStop}%
\bibitem [{\citenamefont {Jordan}\ \emph {et~al.}(2024)\citenamefont {Jordan},
  \citenamefont {van Zuilen}, \citenamefont {Rouillard}, \citenamefont
  {Martins},\ and\ \citenamefont {Lane}}]{jordan2024prebiotic}%
  \BibitemOpen
  \bibfield  {author} {\bibinfo {author} {\bibfnamefont {S.~F.}\ \bibnamefont
  {Jordan}}, \bibinfo {author} {\bibfnamefont {M.~A.}\ \bibnamefont {van
  Zuilen}}, \bibinfo {author} {\bibfnamefont {J.}~\bibnamefont {Rouillard}},
  \bibinfo {author} {\bibfnamefont {Z.}~\bibnamefont {Martins}}, \ and\
  \bibinfo {author} {\bibfnamefont {N.}~\bibnamefont {Lane}},\ }\href@noop {}
  {\bibfield  {journal} {\bibinfo  {journal} {Communications Earth \&
  Environment}\ }\textbf {\bibinfo {volume} {5}},\ \bibinfo {pages} {234}
  (\bibinfo {year} {2024})}\BibitemShut {NoStop}%
\bibitem [{\citenamefont {Matange}\ \emph {et~al.}(2025)\citenamefont
  {Matange}, \citenamefont {Rajaei}, \citenamefont {Capera-Aragones},
  \citenamefont {Costner}, \citenamefont {Robertson}, \citenamefont {Kim},
  \citenamefont {Petrov}, \citenamefont {Bowman}, \citenamefont {Williams},\
  and\ \citenamefont {Frenkel-Pinter}}]{matange2025evolution}%
  \BibitemOpen
  \bibfield  {author} {\bibinfo {author} {\bibfnamefont {K.}~\bibnamefont
  {Matange}}, \bibinfo {author} {\bibfnamefont {V.}~\bibnamefont {Rajaei}},
  \bibinfo {author} {\bibfnamefont {P.}~\bibnamefont {Capera-Aragones}},
  \bibinfo {author} {\bibfnamefont {J.~T.}\ \bibnamefont {Costner}}, \bibinfo
  {author} {\bibfnamefont {A.}~\bibnamefont {Robertson}}, \bibinfo {author}
  {\bibfnamefont {J.~S.}\ \bibnamefont {Kim}}, \bibinfo {author} {\bibfnamefont
  {A.~S.}\ \bibnamefont {Petrov}}, \bibinfo {author} {\bibfnamefont {J.~C.}\
  \bibnamefont {Bowman}}, \bibinfo {author} {\bibfnamefont {L.~D.}\
  \bibnamefont {Williams}}, \ and\ \bibinfo {author} {\bibfnamefont
  {M.}~\bibnamefont {Frenkel-Pinter}},\ }\href@noop {} {\bibfield  {journal}
  {\bibinfo  {journal} {Nature Chemistry}\ ,\ \bibinfo {pages} {1}} (\bibinfo
  {year} {2025})}\BibitemShut {NoStop}%
\bibitem [{\citenamefont {Jennewein}\ \emph {et~al.}(2023)\citenamefont
  {Jennewein}, \citenamefont {Lee}, \citenamefont {Kurtz}, \citenamefont
  {Dizon}, \citenamefont {Shaeffer}, \citenamefont {Chapman}, \citenamefont
  {Chiquete}, \citenamefont {Burks}, \citenamefont {Carlson}, \citenamefont
  {Mason}, \citenamefont {Kobawala}, \citenamefont {Jagadeesan}, \citenamefont
  {Basani}, \citenamefont {Battelle}, \citenamefont {Belshe}, \citenamefont
  {McCaffrey}, \citenamefont {Brazil}, \citenamefont {Inumella}, \citenamefont
  {Kuznia}, \citenamefont {Buzinski}, \citenamefont {Shah}, \citenamefont
  {Dudley}, \citenamefont {Speyer},\ and\ \citenamefont
  {Yalim}}]{jennewein_sol_2023}%
  \BibitemOpen
  \bibfield  {author} {\bibinfo {author} {\bibfnamefont {D.~M.}\ \bibnamefont
  {Jennewein}}, \bibinfo {author} {\bibfnamefont {J.}~\bibnamefont {Lee}},
  \bibinfo {author} {\bibfnamefont {C.}~\bibnamefont {Kurtz}}, \bibinfo
  {author} {\bibfnamefont {W.}~\bibnamefont {Dizon}}, \bibinfo {author}
  {\bibfnamefont {I.}~\bibnamefont {Shaeffer}}, \bibinfo {author}
  {\bibfnamefont {A.}~\bibnamefont {Chapman}}, \bibinfo {author} {\bibfnamefont
  {A.}~\bibnamefont {Chiquete}}, \bibinfo {author} {\bibfnamefont
  {J.}~\bibnamefont {Burks}}, \bibinfo {author} {\bibfnamefont
  {A.}~\bibnamefont {Carlson}}, \bibinfo {author} {\bibfnamefont
  {N.}~\bibnamefont {Mason}}, \bibinfo {author} {\bibfnamefont
  {A.}~\bibnamefont {Kobawala}}, \bibinfo {author} {\bibfnamefont
  {T.}~\bibnamefont {Jagadeesan}}, \bibinfo {author} {\bibfnamefont {P.~B.}\
  \bibnamefont {Basani}}, \bibinfo {author} {\bibfnamefont {T.}~\bibnamefont
  {Battelle}}, \bibinfo {author} {\bibfnamefont {R.}~\bibnamefont {Belshe}},
  \bibinfo {author} {\bibfnamefont {D.}~\bibnamefont {McCaffrey}}, \bibinfo
  {author} {\bibfnamefont {M.}~\bibnamefont {Brazil}}, \bibinfo {author}
  {\bibfnamefont {C.}~\bibnamefont {Inumella}}, \bibinfo {author}
  {\bibfnamefont {K.}~\bibnamefont {Kuznia}}, \bibinfo {author} {\bibfnamefont
  {J.}~\bibnamefont {Buzinski}}, \bibinfo {author} {\bibfnamefont {D.~D.}\
  \bibnamefont {Shah}}, \bibinfo {author} {\bibfnamefont {S.~M.}\ \bibnamefont
  {Dudley}}, \bibinfo {author} {\bibfnamefont {G.}~\bibnamefont {Speyer}}, \
  and\ \bibinfo {author} {\bibfnamefont {J.}~\bibnamefont {Yalim}},\ }in\ \href
  {\doibase 10.1145/3569951.3597573} {\emph {\bibinfo {booktitle} {Practice and
  {{Experience}} in {{Advanced Research Computing}} 2023: {{Computing}} for the
  {{Common Good}}}}},\ \bibinfo {series and number} {{{PEARC}} '23}\ (\bibinfo
  {publisher} {Association for Computing Machinery},\ \bibinfo {address} {New
  York, NY, USA},\ \bibinfo {year} {2023})\ pp.\ \bibinfo {pages}
  {296--301}\BibitemShut {NoStop}%
\end{thebibliography}%
\bibliographystyle{apsrev4-1}

\end{document}